\documentclass[num-refs]{wiley-article}

\usepackage{siunitx}
\usepackage{listings} 
\definecolor{backcolour}{rgb}{0.95,0.95,0.95}
\papertype{Tutorial Reviews}

\title{Response properties of embedded molecules through the polarizable embedding model}

\author[1]{Casper Steinmann}
\author[2]{Peter Reinholdt}
\author[2]{Morten Steen N\o{}rby}
\author[2]{Jacob Kongsted}
\author[2\authfn{1}]{J\'{o}gvan Magnus Haugaard Olsen}

\affil[1]{Department of Chemistry and Bioscience, Aalborg University, Aalborg \O{}, DK-9220, Denmark}
\affil[2]{Department of Physics, Chemistry and Pharmacy, University of Southern Denmark, DK-5230 Odense M, Denmark}

\corraddress{J\'{o}gvan Magnus Haugaard Olsen, Hylleraas Centre for Quantum Molecular Sciences, Department of Chemistry, UiT The Arctic University of Norway, N-9037 Troms\o{}, Norway}
\corremail{jogvan.m.olsen@uit.no}

\presentadd[\authfn{1}]{Hylleraas Centre for Quantum Molecular Sciences, Department of Chemistry, UiT The Arctic University of Norway, N-9037 Troms\o{}, Norway}

\fundinginfo{Danish Council for Independent Research, Grant IDs: DFF--4181-00370 (C.S.) and DFF--7014-00050B (J.K.); H2020-MSCA-ITN-2017 COSINE, Project ID: 765739 (J.K.); Villum Foundation (J.K.); Carlsberg Foundation, Grant ID: CF15-0823 (J.M.H.O)}

\runningauthor{Steinmann et al.}

\begin{document}

\maketitle

\begin{abstract}
The polarizable embedding (PE) model is a fragment-based quantum--classical approach aimed at accurate inclusion of environment effects in quantum-mechanical response property calculations. The aim of this tutorial is to give insight into the practical use of the PE model. Starting from a set of molecular structures and until you arrive at the final property, there are many crucial details to consider in order to obtain trustworthy results in an efficient manner. To lower the threshold for new users wanting to explore the use of the PE model, we describe and discuss important aspects related to its practical use. This includes directions on how to generate input files and how to run a calculation.

\keywords{polarizable embedding, QM/MM, response properties, molecular properties, computational spectroscopy}
\end{abstract}

\section{Introduction}
Hybrid quantum--classical approaches for modeling of chemical or biological systems have in recent years gained considerable interest.
The reason for such popularity of these models relies, to a large degree, on their efficiency and the fact that such models enable calculations on systems of sizes that are otherwise impossible using pure quantum-mechanical methods.
The dielectric continuum models belong to the simplest of the quantum--classical approaches\cite{Cramer1999,Tomasi2005}, and models like the polarizable continuum model (PCM)\cite{miertuvs1981electrostatic,cances1997new} are today implemented in many of the available electronic-structure programs.
In addition, such models are very easy to use: based on a predefined set of atomic radii and the dielectric constant of the solvent, the user can include solvation effects based only on a single calculation.
Only one calculation is needed because the dielectric continuum models implicitly include sampling of solvent configurations.
On the other hand, it is well-known that the dielectric continuum models possess several drawbacks, such as the inability to model the directionality of specific intermolecular interactions like hydrogen bonding or $\pi$--$\pi$ stacking.
Because of this, modeling of environment anisotropies, as found in, e.g., protein matrices is lost.

Another class of quantum--classical approaches consists of discrete models where the atomistic detail of the environment is kept, i.e.\ models based on the concept of combined quantum mechanics and molecular mechanics (QM/MM)\cite{warshel1976theoretical,singh1986combined,field1990combined,senn2009qm}.
Discrete models, compared to the dielectric continuum models, realistically describe the environment, but at an increased level of both complexity and computational requirements.
Regarding the latter point, the increase in computational time is not linked to the discrete nature of the environment as such but rather that sampling of solvent configurations (or protein matrices) now has to be included explicitly.
In many cases, this demand for an explicit account of sampling is met by coupling to molecular dynamics (MD) simulations.
In a sequential manner, the MD simulation is first performed, and from this, a number of snapshots are extracted and used for the following quantum--classical calculations\cite{Coutinho1997}.
This means that the "single calculation" which was required by the dielectric continuum model, is now replaced by "a number of" similar calculations (usually in the order of a few hundred) thereby leading to a significant increase in computational time.
Also, as mentioned above, the complexity of the discrete environmental models are more involved than the dielectric continuum models.
This relies on the fact that in the discrete approach, parameters for each atom in the environment are needed, whereas in the continuum model all of this is reduced to the macroscopic dielectric constant.
Depending on the level of sophistication, these parameters include charges and potentially higher-order multipoles, isotropic or anisotropic polarizabilities, and other types of parameters that account for non-electrostatic interactions such as dispersion and exchange repulsion.
Thus, setting up a quantum--classical calculation based on a discrete environment model is much more involved than using a dielectric continuum description.
On the other hand, the increase in complexity means the atomistic representation of the environment is kept intact.
The discrete models are as such more flexible and can be expected to treat a wide range of environments on the same footing and with similar accuracy.

The polarizable embedding (PE) model\cite{olsen2010,olsen2011} is a fragment-based quantum--classical approach that belongs to the class of discrete environment models.
The parameters describing the effects of the environment are derived by splitting the environment into smaller fragments and then performing quantum-mechanical calculations on each fragment.
By relying on such a first-principles fragment-based approach, the PE model will provide results that are in close agreement with full quantum-mechanical calculations, but at a much reduced computational cost.
The permanent charge distribution of each fragment in the environment is modeled by a multi-center multipole expansion which is usually placed at the position of each atom.
In addition, dipole--dipole polarizabilities are also assigned to each expansion point which introduces an explicit polarization in the environment.
The PE model thus belongs to a class of fully polarizable environmental models where the quantum and classical parts mutually polarize each other.
It has been designed for calculations of molecular response properties, which is a distinctive feature compared to other similar embedding schemes that are based on the induced-dipole polarization model\cite{jensen2003discrete,EFP-excited-states,MMpol,Sderhjelm2009Protein}.
The PE model has thus been formulated within a general quantum-mechanical response-theory framework\cite{olsen1985linear,Thorvaldsen2008,Helgaker2012Recent} meaning that it is capable of addressing a wide variety of spectroscopic properties in addition to calculating excited states and their properties.
Specifically, any property that can be calculated based on linear or nonlinear response functions may include environment effects through the PE model\cite{olsen2010,olsen2011,Steindal2016}, including also resonant-convergent linear and quadratic response\cite{pedersen2014damped,list2014lanczos,Nrby2018}.
London atomic orbitals (LAOs) are supported for magnetic properties that can be obtained from linear response.\cite{Steinmann2014Nuclear}
Finally, the PE model has been combined with a number of electronic-structure methods including Hartree-Fock (HF) and Kohn-Sham density functional theory (DFT)\cite{olsen2010,olsen2011} but also correlated wave-function-based methods such as coupled cluster (CC)\cite{sneskov2011polarizable,schwabe2012peri}, multiconfigurational self-consistent-field theory (MCSCF)\cite{hedegaard2013multi}, multiconfigurational short-range DFT (MC-srDFT)\cite{hedegaard2015polarizable}, the second-order polarization propagator approach (SOPPA)\cite{eriksen2012importance} and also in a relativistic framework using DFT\cite{Hedegrd2017}.
For a recent perspective on the PE model we refer to Ref.~\citenum{List2016Perspective}.
The user is thus able to choose among different electronic-structure methods and select the one that is known to perform best from a cost-efficiency point of view for a given property.
The PE model is per definition not restricted to any specific kind of environment but may be used on the same footing for different environments such as solvents\cite{Beerepoot2016Averaged}, proteins\cite{Olsen2015Accuracy}, DNA\cite{Norby2016Computational}, or lipids\cite{Witzke2017Averaged}.
However, the complexity of setting up calculations may vary depending essentially on the fragmentation procedure used for the environment.
For solvents, the environment is typically fragmented into individual solvent molecules but for large molecules such as proteins the fragmentation becomes more involved because it is necessary to cut covalent bonds in order to define the fragments.

The PE model is implemented in the PE library~\cite{pelib} which has been interfaced to the Dalton\cite{daltonpaper,dalton} and DIRAC\cite{DIRAC17} programs.
We note that the PE model has also been implemented in other electronic structure codes, separately from the implementation in the PE library, including PE-RI-CC2\cite{schwabe2012peri} in TURBOMOLE\cite{turbomole}.
In this tutorial, we will focus on the implementation in the Dalton program.
Most, if not all, of the aspects covered here, will also apply to its use in other programs, albeit with different input.
We will give a quick overview of the theoretical background but expect the reader to already be familiar with the general theoretical concepts of the model.
Instead the focus in this paper is on the practical use of the PE model in response property calculations.
We describe the format of the input files including how they can be generated and provide examples of how to perform calculations in Dalton.
In addition, we will discuss and provide guidelines related to crucial aspects such as the choice of basis set and size of the quantum region, and how the size of the environment and the quality of the embedding potential influence the results of the calculations.
This article can thereby serve as a tutorial for new users that want to explore the use of the PE model for applications related to spectroscopic processes in chemical or biological systems.

\section{Theoretical background}
\label{sec:theory}
The PE model divides the total system into a smaller quantum region, which is treated with a quantum-chemistry method, and its environment, which is represented by multipoles and polarizabilities.
The environment is further split into a number of small fragments, and for each of them, the multipoles and polarizabilities are derived using a multi-center multipole expansion based on quantum-mechanical densities.
If the environment consists of small solvent molecules, such as water, the definition of the fragments is straight-forward.
Larger molecules, such as proteins, require more sophisticated fragmentation schemes.
The fragment calculations are done in isolation, and as such the multipoles and polarizabilities represent the permanent and induced charge density of the fragments in the environment, respectively.
The current implementation of the PE model supports up to fifth-order multipoles which allows highly accurate reproduction of electrostatic potentials (ESPs).
Furthermore, the inclusion of anisotropic dipole--dipole polarizabilities allows efficient calculation of the polarization in the environment.
This includes also mutual polarization interactions between the quantum region and its environment.
The effects of the environment on the quantum region are included by constructing an effective Hamiltonian, which, within HF/DFT methods, results in an effective Fock/Kohn-Sham operator
\begin{equation} \label{eq:effectivefock}
\hat{f}^{\mathrm{eff}}=\hat{f}^{\mathrm{vac}}+\hat{v}^{\mathrm{PE}} \ ,
\end{equation}
where $\hat{f}^{\mathrm{vac}}$ is the "vacuum" Fock/Kohn-Sham operator describing all the internal interactions within the quantum region and $\hat{v}^{\mathrm{PE}}$ is the PE operator describing the potential exerted on the quantum region by the environment, i.e.\ the embedding potential.
The PE operator, in its simplest form, is divided into an electrostatic and an induction (polarization) term
\begin{equation} \label{eq:v_PE}
\hat{v}^{\mathrm{PE}}=\hat{v}^{\mathrm{es}}+\hat{v}^{\mathrm{ind}} \ .
\end{equation}
The electrostatic operator, $\hat{v}^{\mathrm{es}}$, describes the potential from the permanent charge distributions of the environment fragments, i.e.\ the nuclei and (multipole-expanded) electron densities, on the quantum region.
The electrostatic operator can be generally defined as
\begin{equation} \label{eq:v_es}
\hat{v}^{\mathrm{es}}=\sum_{s=1}^{N}\sum_{\left|k\right|=0}^{K}\frac{\left(-1\right)^{\left|k\right|}}{k!}M_{s}^{\left(k\right)}\hat{V}_{s,\mathrm{el}}^{\left(k\right)}
\end{equation}
using a multi-index notation\cite{olsen2014thesis}.
Here, $M_{s}^{\left(k\right)}$ are the multipoles on expansion site $s$ and $\hat{V}_{s,\mathrm{el}}^{\left(k\right)}$ is the operator giving a derivative of the electric potential at site $s$.
Written up to second order this corresponds to
\begin{equation} \label{eq:v_es_simple}
\hat{v}^{\mathrm{es}}=\sum_{s=1}^{N}\left(q_{s}\hat{V}_{s,\mathrm{el}}-\sum_{\alpha}\mu_{s}^{\alpha} \hat{V}_{s,\mathrm{el}}^{\alpha} +\sum_{\alpha,\beta}\Theta_{s}^{\alpha\beta}\hat{V}_{s,\mathrm{el}}^{\alpha\beta}\right) \ ,
\end{equation}
where $\alpha$ and $\beta$ denote Cartesian coordinates and $q$, $\bm{\mu}$, and $\bm{\Theta}$ are the charges, dipoles and quadrupoles, respectively.
The induction operator, $\hat{v}^{\mathrm{ind}}$, describes the effects from the polarized charge distributions, described by induced dipoles, of the environment fragments on the quantum region, and it is defined as
\begin{equation} \label{eq:v_ind}
\hat{v}^{\mathrm{ind}}=-\sum_{s=1}^{N}\bm{\mu}_{s}^{\mathrm{ind}}\left(\mathbf{F}_{\mathrm{tot}}\right)\hat{\mathbf{F}}_{s,\mathrm{el}} \ .
\end{equation}
The induced dipoles $\bm{\mu}_{s}^{\mathrm{ind}}$ are set up by the total electric field at the polarizable site $s$, 
$\mathbf{F}_{\mathrm{tot}}$, i.e.\ the field from the electrons and nuclei in the quantum region, the permanent multipole moments in the environment, as well as other induced dipoles in the environment, and $\hat{\mathbf{F}}_{s,\mathrm{el}}$ is the operator yielding the electronic electric field at a polarizable site $s$.
In order to build the $\hat{v}^{\mathrm{ind}}$ operator, the induced dipoles must first be obtained.
This is done through
\begin{equation} \label{eq:indmom}
\bm{\mathbf{\mu}}_{s}^{\mathrm{ind}}\left(\mathbf{F}_{\mathrm{tot}}\right)=\bm{\alpha}_{s}\mathbf{F}_{\mathrm{tot}}(\mathbf{R}_{s})=\bm{\alpha}_{s}\left(\mathbf{F}_{}(\mathbf{R}_{s})+\sum_{s'\neq s}\mathbf{T}_{ss'}^{(2)}\bm{\mathbf{\mu}}_{s'}^{\mathrm{ind}}\right) \ ,
\end{equation}
where $\mathbf{F}_{}(\mathbf{R}_{s})$ is the electric field at site $s$ from the nuclei, electrons and permanent multipole moments (but not the induced dipoles as this is separately treated in the last term of the above equation), and $\mathbf{T}_{ss'}^{(2)}$ is the so-called dipole--dipole interaction tensor.
Equation~\ref{eq:indmom} leads to a set of coupled equations which can be formulated as a matrix-vector equation, by introducing a column vector containing the induced dipoles $\bm{\mu}^{\mathrm{ind}}=\left(\bm{\mu}_{1}^{\mathrm{ind}},\bm{\mu}_{2}^{\mathrm{ind}},\ldots,\bm{\mu}_{N}^{\mathrm{ind}}\right)^{T}$,
and one containing the electric fields $\mathbf{F}=\left(\mathbf{F}_{}(\mathbf{R}_{1}),\mathbf{F}_{}(\mathbf{R}_{2}),\ldots,\mathbf{F}_{}(\mathbf{R}_{N})\right)^{T}$.
The induced dipole moments are then formally obtained as the solution to the matrix-vector equation
\begin{equation} \label{eq:direct_solver}
\bm{\mu}^{\mathrm{ind}}=\mathbf{BF} \ ,
\end{equation}
where $\mathbf{B}$ is the ($3N \times 3N$) classical response matrix (also known as the \textit{relay} matrix), defined as
\begin{equation} \label{eq:classicalresponsematrix}
\mathbf{B}=\begin{pmatrix}\pmb{\alpha}_{1}^{-1} & -\mathbf{T}_{12}^{(2)} & \ldots & -\mathbf{T}_{1N}^{(2)}\\
-\mathbf{T}_{21}^{(2)} & \pmb{\alpha}_{2}^{-1} & \ddots & \vdots\\
\vdots & \ddots & \ddots & -\mathbf{T}_{(N-1)N}^{(2)}\\
-\mathbf{T}_{N1}^{(2)} & \dots & -\mathbf{T}_{N(N-1)}^{(2)} & \pmb{\alpha}_{N}^{-1}
\end{pmatrix}^{-1} \ .
\end{equation}
The response matrix contains, on the diagonals, the inverse of the polarizabilities, while the off-diagonal elements contain the dipole--dipole interaction tensors.
In practice, the induced dipoles can be obtained by either explicitly constructing the response matrix and directly solving equation~\ref{eq:direct_solver}, or by iterative schemes when an explicit construction is not computationally feasible.
Note that since the induced dipoles depend on the fields exerted by the electron density of the quantum region, which in turn depends on the induced dipoles through the induction operator, it is natural to use a self-consistent scheme, in which the induced dipoles are updated in every SCF cycle, leading to a mutual relaxation of the quantum wave function/density and the induced dipoles of the environment.

One of the main advantages of the PE model is that it is formulated within the general framework of quantum-mechanical response theory.
A review of the theoretical framework behind response theory that include the PE model is beyond the scope of this work and the interested reader is instead referred to the literature indicated in the introduction.
The formulation within response theory allows for the calculation of a wealth of molecular response properties, including among others optical properties like excitation energies, one- and two-photon transition moments, as well as magnetic properties including nuclear shieldings, spin-spin coupling constants and hyperfine coupling tensors.
The presence of an environment is included by replacing the standard (vacuum) Hamiltonian with the effective Hamiltonian.
Furthermore, due to the induction operator, new contributions appear in the response equations.
In the case of linear response, this amounts to additional contributions to the electronic Hessian that describe the response of the environment to the changes in the quantum region induced by the external field.
Another difference between response theory for a molecule in isolation and when embedded into an environment is that the external field is modified by the presence of the polarizable environment.
For optical properties this gives rise to the so-called effective external field effects as described in Ref.~\citenum{List2016Local}.
Note that this leads to changes only in the magnitude of the field, but leaves the frequency unaltered.
As such, the pole structure of the response function is the same, and only quantities such as transition moments are modified, while excitation energies are left unchanged.
Including these effects can be of utmost importance if the aim is to accurately reproduce results from a fully quantum-mechanical calculation.
 
\section{Implementation aspects}
\label{sec:implementation}
\begin{figure}[h!]
	\centering
	\includegraphics[width=0.8\textwidth]{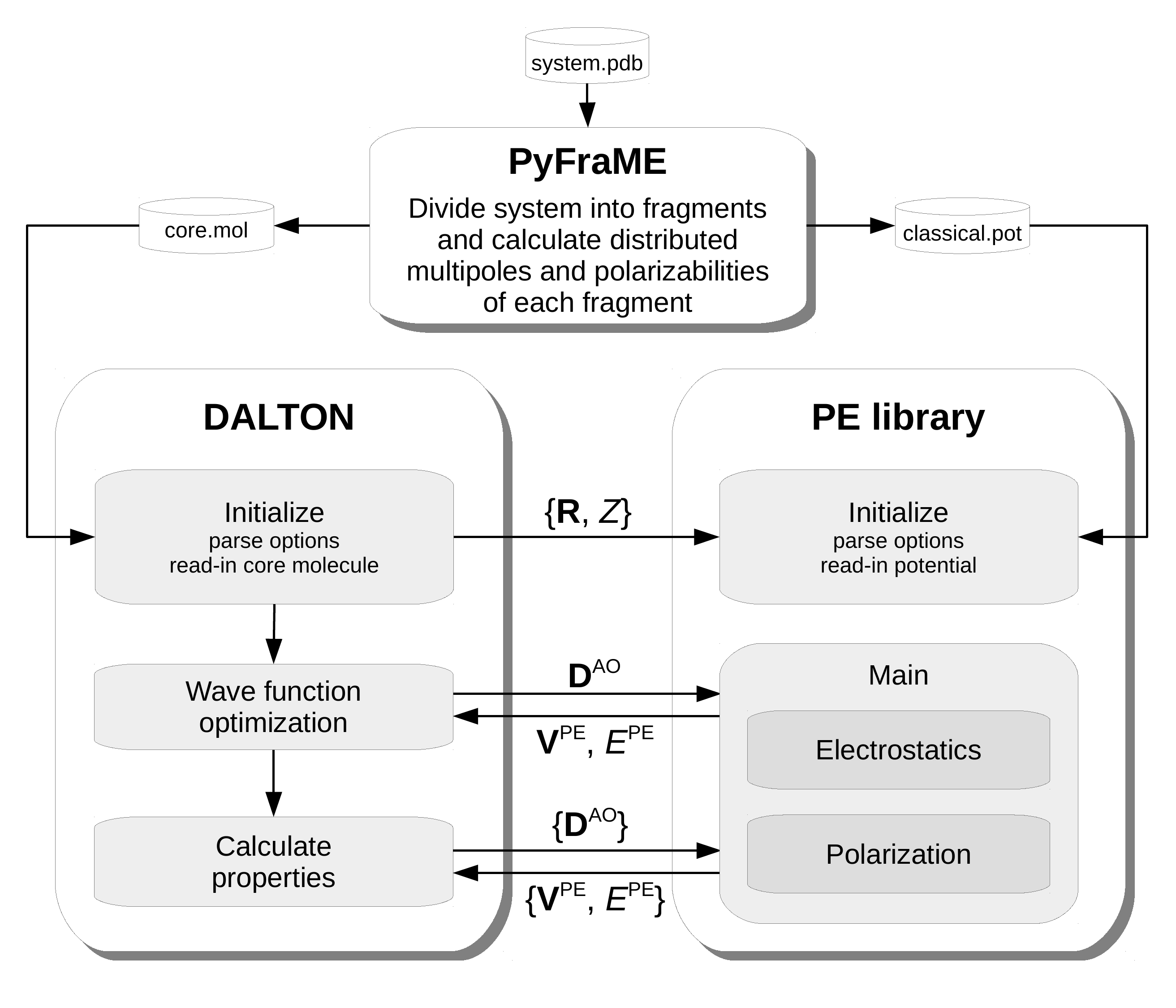}
	\caption{Workflow diagram of a response property calculation. The PE library receives nuclear coordinates and charges of the core quantum region during initialization. The main procedure takes a set of density matrices as input and returns the corresponding Fock matrix contributions and energies.}
	\label{pe_workflow}
\end{figure}

The PE model is implemented in the PE library\cite{pelib} which is written in modern Fortran and is distributed under a GNU LGPLv3 license.
The library has been interfaced to the Dalton\cite{dalton} and DIRAC\cite{DIRAC17} programs.
The only requirements are BLAS and LAPACK as well as the Gen1Int integral package\cite{gen1int,Gao2011}.
Furthermore, the library uses message-passing interface (MPI) version 2.0 standards for a simple master/slave parallelization scheme in which blocks of classical sites are assigned to their own MPI process.
The interface consists of a single main routine apart from the initialization and finalization routines.
The coordinates and charges of the nuclei in the core quantum region are received from the host program in the initialization call.
The initialization also involves parsing options and reading the potential input file (see Section~\ref{ssec:potfile}).
The main routine computes the PE contributions according to a run-type which is specified in each call.
The run-type determines whether the library is required to calculate contributions to the Fock matrix, to the response, or to the magnetic-field or molecular gradients.
The library implements an atomic-orbital (AO) formulation of the PE model which thus makes it completely general with respect to the method used for the quantum region as well as time-independent and time-dependent formalisms.
The main routine takes a number of AO density matrices as input and computes the corresponding Fock matrix contributions and expectation values if relevant.
The fact that it is possible to send in several density matrices makes it possible to loop over density matrices inside the library and thus only calculate one-electron integrals once.
This can give a tremendous speedup especially in response-property calculations.
An illustration of a basic workflow is presented in Figure~\ref{pe_workflow}.

The multipole moments, polarizabilities and interaction tensors consist of $3^{n}$ elements where $n$ is the tensor rank.
However, since they have permutational symmetry, e.g.\ $xyz=yzx$ , we only evaluate the $(n+1)(n+2)/2$ unique components using an ordered array starting from the $x$ components and then increasing from the right (the so-called anti-canonical order), e.g.\ $xxx$, $xxy$, $xxz$, $xyy$, $\ldots$, $zzz$.
The Cartesian components of the interaction tensors are calculated using the open-ended formula by \citeauthor{Dykstra88}\cite{Dykstra88}
\begin{equation} \label{eq:Tk_equation}
T^{(k)} = \sum_{i_{x}=0}^{k_{x}} C^{(1)}_{k_{x}i_{x}} \left(\frac{r_{x}}{|\mathbf{r}|}\right)^{i_{x}} \sum_{i_{y}=0}^{k_{y}} C^{(i_{x}+k_{x}+1)}_{k_{y}i_{y}} \left(\frac{r_{y}}{|\mathbf{r}|}\right)^{i_{y}} \sum_{i_{z}=0}^{k_{z}} C^{(i_{x}+k_{x}+i_{y}+k_{y}+1)}_{k_{z}i_{z}} \left(\frac{r_{z}}{|\mathbf{r}|}\right)^{i_{z}} \frac{1}{|\mathbf{r}|^{|k|+1}} \ .
\end{equation}
The $C^{(n)}_{ij}$ coefficient array is generated during the initialization and stored in-memory whereas the components of the interaction tensors are calculated on-the-fly when needed.
The induced dipole moments can be determined either using a direct algorithm (utilizing BLAS/LAPACK) or iteratively using the Gauss-Seidel method (also known as the method of successive displacement)
\begin{equation} \label{eq:iterative_solver}
\bm{\mu}^{[i]}_{s}=\bm{\alpha}_{s}\left(\mathbf{F}(\mathbf{R}_{s}) + \sum_{s` < s} \mathbf{T}^{(2)}_{s`s} \bm{\mu}^{[i]}_{s`} + \sum_{s` > s} \mathbf{T}^{(2)}_{s`s} \bm{\mu}^{[i-1]}_{s`}\right) \ ,
\end{equation}
Here the $i$ index refers to the current iteration such that $\bm{\mu}^{[i-1]}_{s}$ is an induced dipole from the previous iteration.
In a parallel calculation, however, it is a combined Jacobi and Gauss-Seidel iterative scheme, i.e.\ within a block of sites given to a process the induced moments are determined using Gauss-Seidel but between each block it corresponds to Jacobi iterations.
By default, the induced dipoles are converged using a threshold defined as $\sum_{s=1}|\bm{\mu}_s^{[i]} - \bm{\mu}_s^{[i-1]}| < 10^{-8}$.
 
\section{Practical Aspects of Polarizable Embedding}
In this part of the tutorial, we describe and discuss crucial aspects related to response property calculations using the PE model.
First, we provide a getting-started guide where we show how to generate the input files for Dalton and the PE library, we describe the format of the input files, and also demonstrate how to run the calculations.
We will then discuss how different key aspects can impact the results of response property calculations.

\subsection{Getting Started}
\label{sec:general}
Performing a PE-QM calculation in Dalton requires three input files: the molecule file (with default file extension \emph{.mol}), the potential file (\emph{.pot}), and the Dalton input file (\emph{.dal}).
The molecule file defines the quantum region including basis set information, while the potential file contains the embedding parameters, which includes coordinates, multipole moments (e.g.\ $q$, $\bm{\mu}$, $\bm{\Theta}$, etc.) and polarizabilities ($\bm{\alpha}$) of all classical sites.
The input file specifies the type of calculation to perform, which includes the type of wave function and property as well as options related to the PE model.
We will demonstrate how the molecule and potential files can be generated and briefly describe the basic format of the three input files.
We refer to the Dalton manual for a more detailed description of the Dalton input and molecule file (\url{http://www.daltonprogram.org/}).
We conclude with some examples that show how to run PE-QM calculations in practice.
For this you will need the Dalton program (see \url{http://www.daltonprogram.org/} for details on how to download and install Dalton).
Note that some of the features presented in this tutorial may only be available in a development version of Dalton.
The development version and the last few releases can be downloaded from the public source code repository (\url{https://gitlab.com/dalton/dalton}).
The files used in the examples below are available at \url{https://doi.org/10.5281/zenodo.1212345} together with Dalton's PE-QM tests which can be used as inspiration for further exploration.

\subsubsection{Generating Input Files}
\label{ssec:generatefiles}
It can be quite cumbersome to manually generate embedding potentials for a large set of molecules, particularly for complex systems such as proteins which require a more advanced fragmentation scheme.
For this reason, several tools have been developed but here we provide examples using the PyFraME\cite{pyframe} Python package which can automatically generate potentials given an input structure.
The simplest way to download and install PyFraME is from the Python Package Index (PyPI) where the installation can be performed simply by issuing the command \texttt{pip install pyframe} in your shell environment.
The development repository can be found at \url{https://gitlab.com/FraME-projects/PyFraME} and all release versions are also deposited at \url{https://doi.org/10.5281/zenodo.775113}.

The basic workflow consists of three main steps.
First a \texttt{MolecularSystem} object is created which requires a PDB input file containing the molecular structure.
Fragments are then extracted from the system and placed in a region that is treated according to specified input arguments.
Any number of regions may be added making it easy to use different parameters for different parts of the environment.
In fact, the settings for each region are highly customizable but reasonable defaults will be used unless otherwise specified.
Finally, a \texttt{Project} object is created which handles all computational aspects using information from a \texttt{MolecularSystem} object, i.e.\ creating the embedding potential, possibly involving a number of fragment calculations, and writing the input files which can then be used in further calculations.
Listing~\ref{lst:pyframe_sep} illustrates the workflow using a simple example.
It shows how to create the molecule and potential file for a system that consists of an acrolein molecule embedded in a small droplet of water.
The acrolein molecule is extracted by providing the name as it is specified in the PDB input file.
It is placed in the core (quantum) region and treated using Jensen's segmented contracted\cite{Jensen2014Unifying} polarization consistent\cite{Jensen2001Polarization,Jensen2008Basis} basis set (pcseg-1).
The solvent molecules are selected according to their distance from the quantum region but can also be chosen according to other parameters, such as name or chain id.
By default the distances are calculated using the center of mass of the fragments, so in this example all water molecules whose center of mass is within 4.0 \AA{} from the center of mass of acrolein are selected.
For demonstration purposes, we use a short distance here, which will only extract two water molecules, whereas real applications would contain many more solvent molecules.
It is also possible to use a nearest atom distance criterion by adding \texttt{use\_center\_of\_mass=False} as an argument.
The solvent is added to a region which is defined to use a built-in standard potential termed solvent embedding potential (SEP)\cite{Beerepoot2016Averaged}.
After the partitioning of the system into the core region and the solvent environment, the embedding potential is created.
This process may include a series of calculations but in this example we use a standard potential so it is quickly completed.
The total formal charge as provided in the PDB input file and the sum of partial charges that enter in the embedding potential are printed after the embedding potential is created.
Any surplus charge will be redistributed to all classical sites but beware that a significant discrepancy is a strong indication that something is wrong.
Finally, a molecule and potential file containing the acrolein molecule and embedding parameters of the solvent, respectively, are written to disk.
This procedure will create a new directory containing the molecule and potential files which are all named after the input file.
In the following sections, we will describe the format of the files and will furthermore also use them in example calculations.

\begin{lstlisting}[language=Python, label=lst:pyframe_sep, caption=Using the PyFraME package to generate a molecule and potential file for a solute-solvent system.]
>>> import pyframe
>>> system = pyframe.MolecularSystem(input_file='acrolein_water.pdb')
>>> core = system.get_fragments_by_name(names=['ACR'])
>>> system.set_core_region(fragments=core, program='Dalton', basis='pcseg-1')
>>> solvent = system.get_fragments_by_distance(reference=core, distance=4.0)
>>> system.add_region(name='solvent', fragments=solvent, use_standard_potentials=True,
>>>                   standard_potential_model='SEP')
>>> project = pyframe.Project()
>>> project.create_embedding_potential(system)
INFO: total formal charge:   0.0000
INFO: sum of partial charges:   0.0000
>>> project.write_core(system)
>>> project.write_potential(system)
\end{lstlisting}

The example in Listing~\ref{lst:pyframe_sep} used a standard potential included in PyFraME that consists of isotropic parameters (charges and isotropic polarizabilities) which is available for several simple solvents.
These parameters offer acceptable quality for a wide range of purposes\cite{Beerepoot2016Averaged}.
For calculations with higher demands on the accuracy, or in cases where embedding parameters are not available, PyFraME has the ability to derive parameters directly from quantum-mechanical calculations.
In Listing~\ref{lst:pyframe_protein}, we show an example that will calculate the embedding parameters of a protein.
In this example, the fragments are retrieved by specifying the chain id's as they are given in the PDB input file.
They are added to a region where bonded fragments are treated using the molecular fractionation with conjugate caps (MFCC) scheme\cite{Zhang2003,Soderhjelm2009}.
The parameter set will consist of multipoles up to second order (i.e.\ quadrupoles) and dipole-dipole polarizabilities, both of which are obtained using the localized properties (LoProp) model\cite{Gagliardi2004Local}.
By default PyFraME will use Dalton and the LoProp for Dalton script\cite{Vahtras2014LoProp} for the individual fragment calculations.
Note also that the LoProp procedure requires an ANO-type\cite{Almlof1991ANO} basis set\cite{Gagliardi2004Local}.
Several commonly used basis sets have been recontracted to ANO-type and are available in Dalton with a \texttt{loprop-} prefix.
PyFraME handles the execution and processing of the many smaller sub-tasks involved in obtaining the embedding parameters.
The smaller tasks can be spread across multiple processes in parallel on a single node, as show in the example, and it is further possible to run across multiple nodes on clusters.
The settings for the calculations can be modified either through arguments when creating the Project object or by setting the variables after creation as demonstrated in the example.
PyFraME will attempt to retrieve the list of nodes if the script is running on a cluster with a job scheduler (currently PBS and SLURM are supported).
Note that default settings will be used so it is always a good idea check the settings by printing them as shown in the example.

\begin{lstlisting}[language=Python, label=lst:pyframe_protein, caption=Using the PyFraME package to generate a potential representing the insulin protein.]
>>> import pyframe
>>> system = pyframe.MolecularSystem('/path/to/work/insulin.pdb')
>>> protein = system.get_fragments_by_chain_id(chain_ids=['A', 'B'])
>>> system.add_region(name='protein', fragments=protein, use_mfcc=True,
>>>                   use_multipoles=True, multipole_order=2, multipole_model='LoProp',
>>>                   multipole_method='DFT', multipole_xcfun='PBE0',
>>>                   multipole_basis='loprop-6-31+G*', use_polarizabilities=True,
>>>                   polarizability_model='LoProp', polarizability_method='DFT',
>>>                   polarizability_xcfun='PBE0', polarizability_basis='loprop-6-31+G*')
>>> project = pyframe.Project(jobs_per_node=24)
>>> project.scratch_dir = /path/to/scratch
>>> project.print_info()
INFO: work directory set to /path/to/work
INFO: scratch directory set to /path/to/scratch
INFO: running 24 job per node
INFO: number of MPI processes per job set to 1
INFO: number of OpenMP threads per job set to 1
INFO: memory per job set to 2048 MB
INFO: memory per MPI process set to 2048.0 MB
INFO: communication port set to 666
INFO: using node(s): ['the_beast']
>>> project.create_embedding_potential(system)
>>> project.write_potential(system)
\end{lstlisting}

\subsubsection{The Molecule File}
\label{ssec:molfile}
The molecule input file for acrolein that was generated in the previous section is shown in Listing~\ref{lst:acrolein}.
The format of the file follows exactly that of a normal Dalton calculation without any environment effects.
The molecule file contains the charge and coordinates of the atoms that define the quantum region.
The coordinates in the example are given in \aa{}ngstr\"{o}m as indicated by the \texttt{Angstrom} keyword (the default unit is bohr).
The basis set is also specified in the molecule file, which in this case is the pcseg-1 basis set.
It is possible to use different basis set for different atom-types.
In such cases the two first lines are replaced by a single line containing the \texttt{AtomBasis} keyword and each line specifying the atom-type is appended with the \texttt{Basis} keyword, e.g.\ \texttt{Charge=8.0 Atoms=1 Basis=pcseg-2}.
The total charge of the quantum region should also be specified, which is set to zero in the example (\texttt{Charge=0}), and note also that point-group symmetry is not supported by the PE library (i.e.\ \texttt{NoSymmetry} keyword).
For a more detailed description of the format of the molecule file, we refer to the Dalton manual which can be found at \url{http://www.daltonprogram.org/}.

\begin{lstlisting}[language=, label={lst:acrolein}, caption=Dalton molecule file with structure of acrolein using the pcseg-1 basis set.]
BASIS
pcseg-1
Core region
Generated by PyFraME 0.1.1
AtomTypes=3 Charge=0 NoSymmetry Angstrom
Charge=8.0 Atoms=1
O      1.542000     0.710000    -0.255000
Charge=6.0 Atoms=3
C      0.674000    -0.126000    -0.067000
C     -0.749000     0.181000     0.071000
C     -1.633000    -0.798000     0.275000
Charge=1.0 Atoms=4
H      0.937000    -1.200000     0.010000
H     -1.044000     1.225000     0.000000
H     -1.313000    -1.832000     0.342000
H     -2.692000    -0.601000     0.381000
\end{lstlisting}

\subsubsection{The Potential File}
\label{ssec:potfile}
The potential file represents the environment and thus contains all the coordinates, multipoles, and polarizabilities associated with the fragments and molecules in the environment.
As an example, we present the potential file that was generated in Section~\ref{ssec:generatefiles} in Listing~\ref{lst:potwat}.
The coordinates are given in the \texttt{@COORDINATES} section which follows the XYZ file format.
Thus, the first line specifies the number $N$ of coordinates followed by a title line specifying the unit which must be either \texttt{AA} or \texttt{AU} corresponding to \aa{}ngstr\"{o}m or bohr, respectively.
There is an implicit indexing of the coordinates starting from 1 to $N$ which is used in the subsequent sections.
The indexes are shown the last column of the coordinate section.
Note, however, that these indexes are not read and are only shown in the example for convenience.
The permanent multipoles are specified in the \texttt{@MULTIPOLES} section.
Here, the multipole order is specified first (\texttt{ORDER 0} for charges, \texttt{ORDER 1} for dipoles and so on) followed by the number of multipoles, $N$, of that particular order to read.
Hereafter follows $N$ lines listing the multipoles of the specified order.
The first value of each line specifies the index of the site where the multipole is located corresponding to the implicit indexing in the coordinate section.
Each line also contains the Cartesian elements of the multipole (or charge in the case of zeroth order such as in the example).
Only the symmetry-unique elements are provided and should be given using the anti-canonical ordering, e.g.\ $xx$, $xy$, $xz$, $yy$, $yz$, and $zz$.
The polarizabilities are listed in the \texttt{@POLARIZABILITIES} section.
Currently, only dipole--dipole polarizabilities are implemented which are defined by the line \texttt{ORDER 1 1}.
They are anisotropic by default and are also specified in anti-canonical order.
It is also possible to use isotropic dipole--dipole polarizabilities, $\alpha_{\mathrm{iso}}$, by replacing the $xx$, $yy$, and $zz$ elements with $\alpha_{\mathrm{iso}}$ and zeroing the $xy$, $xz$, and $yz$ elements, as shown in Listing~\ref{lst:potwat}.
Alternatively, anisotropic polarizabilities can be made isotropic through a keyword (\texttt{.ISOPOL}) in the Dalton input file.
However, use of this keyword will not significantly affect the computational time since they are treated as anisotropic by the PE library.
After \texttt{ORDER 1 1}, the number of polarizabilities, $N$, is given and followed by $N$ lines of polarizabilities with the first value of each line indicating the index of the site.
Following the polarizabilities, are the exclusion lists marked by the \texttt{EXCLISTS} keyword.
The first line after the \texttt{EXCLISTS} keyword has two numbers.
They are the number of exclusion lists $N$ and maximum length $M$ of the lists, respectively.
Hereafter follows $N$ lines of exclusion lists with $M$ site indices.
The first index indicates the site to which the exclusion list belongs to.
This site is not allowed to be polarized by the sites indicated by the remaining indices, e.g.\ a line \texttt{1 2 3} means that the site with index 1 cannot be polarized by electric fields coming from sites 2 and 3.
Note that all lines must have $M$ indices and may therefore need to be padded with zeros.

\begin{lstlisting}[language=, label={lst:potwat}, caption=Potential file containing two water molecules.]
@COORDINATES
6
AA
O     1.56300000    -0.66900000     2.76300000      1
H     2.48300000    -0.41100000     2.70100000      2
H     1.56400000    -1.60700000     2.57300000      3
O    -2.50200000    -0.73500000    -2.85900000      4
H    -2.41400000    -1.00400000    -3.77300000      5
H    -2.13800000     0.15000000    -2.83400000      6
@MULTIPOLES
ORDER 0
6
1    -0.67444000
2     0.33722000
3     0.33722000
4    -0.67444000
5     0.33722000
6     0.33722000
@POLARIZABILITIES
ORDER 1 1
6
1     5.73935000    0.00000000    0.00000000    5.73935000    0.00000000    5.73935000
2     2.30839000    0.00000000    0.00000000    2.30839000    0.00000000    2.30839000
3     2.30839000    0.00000000    0.00000000    2.30839000    0.00000000    2.30839000
4     5.73935000    0.00000000    0.00000000    5.73935000    0.00000000    5.73935000
5     2.30839000    0.00000000    0.00000000    2.30839000    0.00000000    2.30839000
6     2.30839000    0.00000000    0.00000000    2.30839000    0.00000000    2.30839000
EXCLISTS
6 3
1   2  3
2   1  3
3   1  2
4   5  6
5   4  6
6   4  5
\end{lstlisting}

\subsubsection{The Dalton File}
\label{ssec:dalfile}
A basic Dalton input file for a PE-HF ground-state calculation is presented in Listing~\ref{lst:daliter}.
The presence of the \texttt{.PEQM} keyword in the \texttt{**DALTON INPUT} section activates the PE library using default settings.

\begin{lstlisting}[language=, label={lst:daliter}, caption=Dalton input file for a PE-HF ground-state calculation with default settings.]
**DALTON INPUT
.RUN WAVE FUNCTIONS
.PEQM
**WAVE FUNCTIONS
.HF
**END OF DALTON INPUT
\end{lstlisting}

Further options concerning the embedding part of the calculation are controlled through additional keywords under the \texttt{*PEQM} section which is added under the main \texttt{**DALTON INPUT} section.
For a full list of options we refer to the Dalton manual (\url{http://daltonprogram.org/}).
An example that includes the \texttt{*PEQM} section is shown in Listing~\ref{lst:daldirect}.
Here the \texttt{.DIRECT} keyword specifies that induced dipoles should be determined using a direct solver whereas the default is to use an iterative solver (see eq~\ref{eq:iterative_solver}).
However, it is important to note that this requires explicit construction of the $3N\times3N$ classical response matrix (see eq~\ref{eq:classicalresponsematrix}), which is only feasible for rather small systems, where memory requirements are low.
Several other keywords will be discussed throughout the tutorial.
These keywords must be added under the \texttt{*PEQM} section in order to generate the proper Dalton input file.

\begin{lstlisting}[language=, label={lst:daldirect}, caption=Dalton input file for a PE-HF ground-state calculation using the direct solver for induced dipoles.]
**DALTON INPUT
.RUN WAVE FUNCTIONS
.PEQM
*PEQM
.DIRECT
**WAVE FUNCTIONS
.HF
**END OF DALTON INPUT
\end{lstlisting}

\subsubsection{Running Calculations}
\label{ssec:gscalc}
In the following, we will use the Dalton input file shown in Listing~\ref{lst:daliter} and name it \texttt{pe-hf.dal}, the molecule file in Listing~\ref{lst:acrolein} naming it \texttt{acrolein.mol}, and the potential file in Listing~\ref{lst:potwat} naming it \texttt{water.pot}.
The latter two files can be generated as shown in Listing~\ref{lst:pyframe_sep} but all files can also be downloaded at \url{https://doi.org/10.5281/zenodo.1212345}.
Using these input files, Dalton will perform a PE-HF wave function calculation on acrolein with two classical water molecules.
The calculation is carried out by executing the command shown in Listing~\ref{lst:rundalton1}.
The \texttt{-t} option is used to specify a path to a directory where Dalton can place temporary files which are deleted after a successful run.
Alternatively, the path can be specified by setting the \texttt{DALTON\_TMPDIR} environment variable, e.g.\ \texttt{export DALTON\_TMPDIR=/path/to/scratch}.
When the calculation is finished, the output is returned in a file named \texttt{pe-hf\_acrolein\_water.out}.
The default is to name the output file according to the input filenames but this behavior can be changed using the \texttt{-o \textit{filename.out}} option.
More details about this and other options can be found in the Dalton manual (\url{https://daltonprogram.org/}).

\begin{lstlisting}[language=bash, label={lst:rundalton1}, caption=Shell command to run a Dalton calculation.]
$ dalton -t /scratch -dal pe-hf.dal -mol acrolein.mol -pot water.pot
\end{lstlisting}

The PE library prints information regarding the embedding potential and the settings used in the calculation in the Dalton output file which should always be verified.
The PE library also prints the individual PE contributions to the total energy as shown in Listing~\ref{lst:outdalton1}.
The total PE energy is divided into electrostatic (interactions with permanent multipoles) and polarization (interactions with induced dipoles) contributions, and each of those is further divided into electronic, nuclear and multipole contributions.
The energy contributions may be useful to help diagnose problems or instabilities with calculations, in particular in relation to over-polarization effects.
There are different ways to circumvent this problem depending on the origin which will be discussed later in this tutorial.
An important observation is that the absolute values of the nuclear and electronic contributions to the electrostatic and polarization energies should not differ too much.
Large deviations from this observation might be a sign of over-polarization or electron spill-out effects, i.e.\ the "leaking" of electron density into the environment.

\begin{lstlisting}[language=, label={lst:outdalton1}, caption=Final energies from the PE library in the Dalton output.]
    .--------------------------------------------------.
    | Final results from polarizable embedding library |
    '--------------------------------------------------'

        Polarizable embedding energy contributions:
       ---------------------------------------------

       Electrostatic contributions:
            Electronic                 0.333481570904
            Nuclear                   -0.336110988573
            Multipole                  0.000072195000
            Total                     -0.002557222670

       Polarization contributions:
            Electronic                 0.004141454501
            Nuclear                   -0.004296809245
            Multipole                  0.000008474472
            Total                     -0.000146880272

       Total PE energy:            -0.002704102942
\end{lstlisting}

The computational cost of PE-QM calculations are of course higher compared to a calculation in vacuum.
The most costly PE contribution is the polarization part since it requires the determination of induced dipoles every time the electron density is updated (see eq~\ref{eq:v_ind}).
For very large systems, the iterative determination of induced dipoles can become a bottleneck.
In such cases, the computational costs can be reduced by neglecting the many-body induction (polarization) effects in the environment.
This corresponds to the removal of the electric field from other induced dipoles (see eq~\ref{eq:indmom}), i.e.\ there are no interactions between induced dipoles, which thus avoids the need for iterations since the induced dipoles can be determined by a single matrix-vector product.
Such a calculation can be enabled by adding the \texttt{.NOMB} keyword to the \texttt{*PEQM} section in the Dalton input file.
Note that the energy difference is not a good indicator of the severity of the approximation.
It is always advisable to check directly the effects on the property of interest.

The \texttt{.NOMB} keyword can also help identify excessive polarization in the environment that can arise, for example, when two polarizable sites are in too close proximity which can cause a so-called polarization catastrophe.
In such cases, there will be very large induced dipoles which will also trigger warnings in the Dalton output file.
To help remedy situations where there are too large induced dipoles, the PE library includes the option to apply Thole-type exponential damping\cite{Swart2006DRF,Thole1981Molecular} of the electric fields at the polarizable sites.
This includes damping of the fields from induced dipoles (\texttt{.DAMP INDUCED}), from permanent multipoles (\texttt{.DAMP MULTIPOLE}), and from the quantum region (\texttt{.DAMP CORE}).
It is advisable, however, to always investigate the origin of the issue because it can be a sign of poor-quality structures.
This could for example be the case when snapshots are extracted directly from classical MD simulations.
In certain circumstances, e.g.\ when the structural issues are not likely to affect the final results or when using atomic polarizabilities where bonded atoms can polarize each other, damping can remove artifacts related to over-polarization.

It might be beneficial to carry out calculations in multiple steps, for example if multiple response properties are to be determined based on the same ground-state wave function.
Instead of running the same ground-state calculation twice, one approach could be to (1) run a ground-state calculation and (2) run two independent property calculations using the same ground-state wave function.
Restarting from a previous calculation requires the restart files, which are returned in a gzip'ed tar file with the same name as the output file, and the addition of a \texttt{.RESTART} keyword in both the \texttt{*PEQM} and \texttt{**WAVE FUNCTION} sections.
The command to restart Dalton is shown in Listing~\ref{lst:rundalton2} where it is assumed that the DALTON\_TMPDIR environment variable has been set.

\begin{lstlisting}[language=bash, label={lst:rundalton2}, caption=Shell command to run Dalton using restart files from a previous calculation.]
$ dalton -f pe-hf_acrolein_water.tar.gz -dal restart.dal -mol acrolein.mol -pot water.pot
\end{lstlisting}

\subsection{Size of the Quantum Region}
\label{sec:qmregionsize}
One of the crucial aspects of any quantum--classical calculation is the size of the quantum region.
Because of the computational complexity associated with quantum-chemistry methods, i.e.\ the steep scaling with system size and/or large prefactor, the size of the quantum region will often be a compromise between accuracy and computational effort.
The aim is thus to include as few atoms as possible while still being able to reproduce the property of interest to a desired accuracy.
It is of course necessary to include the parts of the system that are directly involved in the property of interest, i.e.\ where the interactions cannot be described adequately by classical electrostatics.
For some types of properties it is apparent which parts should be in the quantum region while for others it can be more difficult to determine.
For example, for an excitation energy and oscillator strength of a well-localized transition in a chromophore, it is natural to include only the chromophore in the quantum region since the involved orbitals will be confined to it.
If we were interested in multi-photon absorption strengths of the same transition, the situation would become more complicated, since other orbitals than those directly involved in the transition may have significant contributions.
This can be especially tricky for chromophores that are covalently bonded to a larger structure such as a protein\cite{Steindal2016}.

For solute-solvent systems in which the dominant interaction between the quantum and classic subsystems is electrostatic, keeping only the solute in the quantum region, and the solvent molecules in the classic region, will usually give acceptable results. 
If, on the other hand, there are strong specific interactions that are not well-described by classical electrostatics, including some of the nearest solvent molecules in the quantum region might become necessary.
In the case of a chromophore that is covalently bonded to a large protein it becomes necessary to cut bonds.
This will unavoidably introduce artifacts in the system whose effects should be minimized.
It is therefore preferable to cut bonds that only introduce minimal perturbation in the system which usually means only cutting aliphatic C--C bonds.
Furthermore, it is also generally desirable to cut bonds as far away as possible from the chromophore.
See Section~\ref{sec:qmmminterface} for more details concerning quantum-classical interfaces.

\begin{figure}[h!]
\centering
\includegraphics[width=0.8\textwidth]{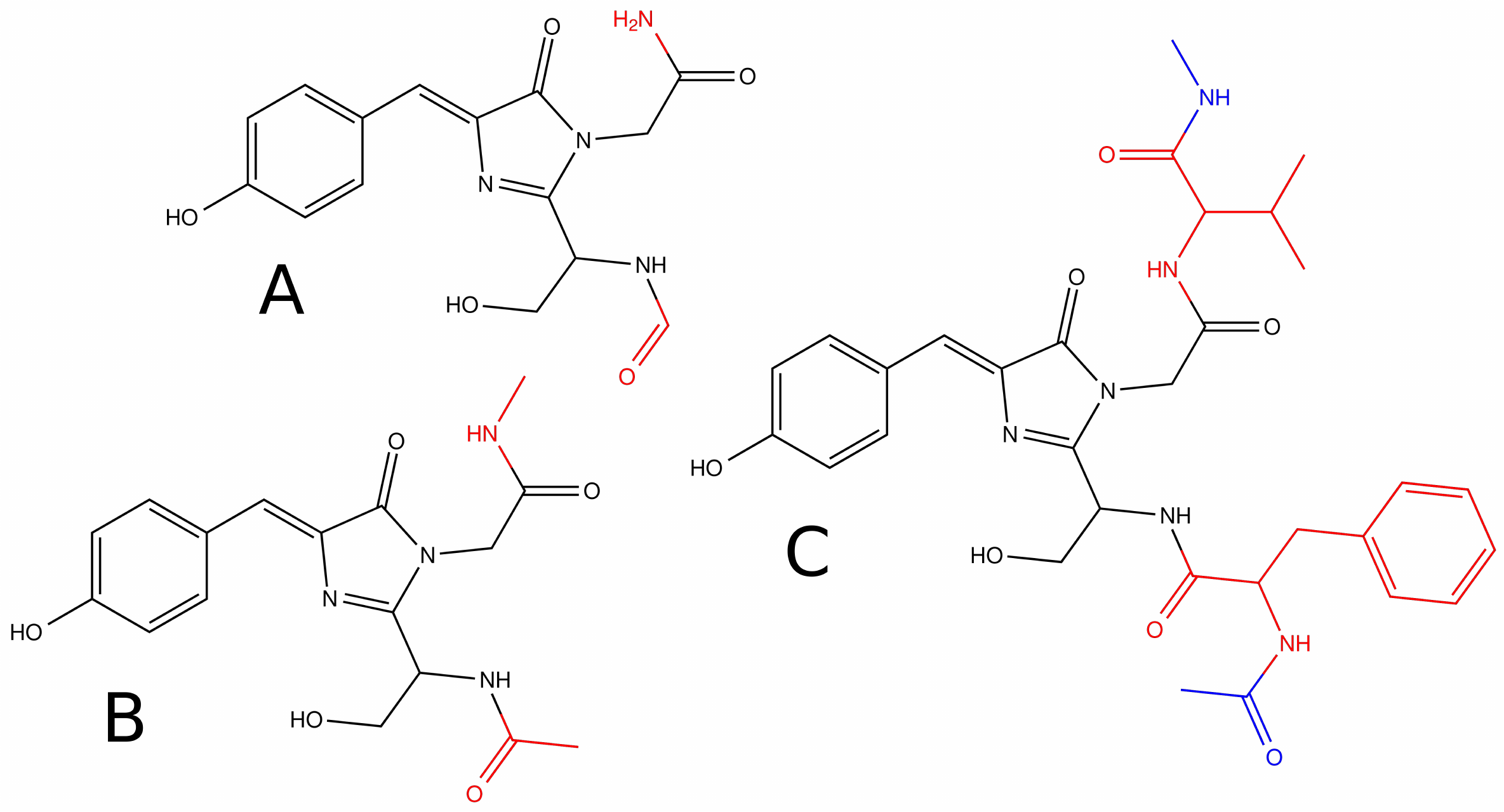}
\caption{Examples of different choices for the quantum region that include the 4-hydroxybenzylidene-1,2-dimethylimidazolinone (HBDI) chromophore and can be used in calculations on the green fluorescent protein (GFP).}
\label{fgr:quantumregion}
\end{figure}

In cases where previous experience or chemical intuition does not provide an answer on how large the quantum region should be, the best strategy is to perform a convergence analysis.
It is usually necessary to analyze the convergence of the specific property that is of interest since the convergence behavior can be different for different properties.
The strategy for increasing the quantum region depends on the system.
For solute-solvent systems it is usually done by including solvent molecules according to their distance to the solute since these are assumed to have the most important interactions.
It is more complicated for chromophores in proteins where a sensible termination of the quantum region is desirable in order to minimize potential artifacts due to the cutting of bonds.
The situation is illustrated in Figure~\ref{fgr:quantumregion} using the 4-hydroxybenzylidene-1,2-dimethylimidazolinone (HBDI) chromophore in the green fluorescent protein (GFP).
In Figure~\ref{fgr:quantumregion}A only the residues that make up the chromophore are included as well as a minimal number of atoms from the neighboring residues which still preserves the amide bonds although they are reduced to primary amides.
The larger region shown in Figure~\ref{fgr:quantumregion}B fully includes the amide bonds by including a larger part of the neighboring residues while the largest region shown in Figure~\ref{fgr:quantumregion}C includes the complete neighboring residues and preserves the amides of their bonded residues.
It may also be appropriate to consider fragments that are not in the immediate vicinity in case there are important interactions, such as charge-transfer or dispersion.
We note that care must be taken for larger quantum regions which in some cases can be problematic when using approximate exchange-correlation functionals\cite{Jakobsen2013,Isborn2013The}.
As such, if large quantum regions are needed, using either a wave-function-based approach or using range-separated functionals with the correct asymptotic limit is strongly recommended.

Many QM/MM studies have shown that a large quantum region is required (typically 500 atoms or more) to ensure convergence with respect to, e.g.\ energies\cite{Sumowski2009,Fox2011,Liao2013,Sadeghian2014,Kulik2016}, forces\cite{Solt2009}, absorption spectra\cite{Isborn2012Electronic,Milanese2017Convergence}, and NMR shielding constants\cite{Flaig2012Convergence}.
However, some of these studies use embedding potentials that are based on standard force fields that have been parametrized for other purposes, for example to reproduce bulk properties of water (e.g.\ TIP3P\cite{Jorgensen1981Quantum,Jorgensen1983Comparison}), and not the electrostatic potential (ESP), which is the key factor in obtaining good accuracy when calculating response properties.
We have shown in a number of studies that the size of the quantum region can be substantially reduced if a high-quality embedding potential is used, as exemplified in the case of excitation energies\cite{Schwabe2011Solvation,Schwabe2015Analysis,Naabo2017Quality}, oscillator strengths\cite{List2016Local,Naabo2017Quality}, and NMR shielding constants\cite{Steinmann2014Nuclear,Steinmann2017Automated}.
See Section~\ref{sec:embedding} for more details concerning the quality of the embedding potential.

\begin{table}[h!]
\centering
\caption{Convergence of excitation energies (in eV) with respect to number of quantum atoms (\textit{N}$_\text{QM}$). Results extracted from Ref.~\citenum{Naabo2017Quality}.}
\label{tbl:devconv}
\begin{threeparttable}
    \begin{tabular}{ccccc}
    \headrow
    \thead{QM$n$} & \thead{\textit{N}$_\text{QM}$} & \thead{QM} & \thead{QM/MM} & \thead{PE-QM} \\
    \hline
    1 & 35  & 3.12 & 3.12  & 2.98  \\
    2 & 96  & 3.11 & 3.15  & 3.02  \\
    3 & 117 & 3.05 & 3.09  & 2.99  \\
    4 & 161 & 3.04 & 3.04  & 2.97  \\
    5 & 281 & 3.00 & 3.00  & 3.00  \\
    6 & 317 & 2.99 & 3.00  & 2.98  \\
    \hline
    \end{tabular}
\end{threeparttable}
\end{table}

As an illustrative example, we highlight some results from a publication by \citeauthor{Naabo2017Quality}\cite{Naabo2017Quality}
In that study, the convergence in terms of quantum-region size was investigated for the lowest intense excitation energy and the associated oscillator strength of a solvated GFP.
The results are shown in Table~\ref{tbl:devconv}.
Only the chromophore is included in the smallest quantum region (QM1) with 35 atoms whereas there are 317 atoms in the largest quantum region (QM6).
In all cases, the quantum region was treated using the CAM-B3LYP\cite{Yanai2004NewHybrid} functional and the 6-31G* basis set\cite{Hehre1972Self,Hariharan1973Influence}.
Point charges from AMBER ff99 were used in the QM/MM calculations while the multipoles and polarizabilities used in PE-QM were based on fragment calculations at B3LYP/6-31G* level.
Included are also results from cluster calculations (QM in Table~\ref{tbl:devconv}) where the environment is ignored completely so that only the quantum region is modeled.
Using electrostatic embedding with only point charges (QM/MM in Table~\ref{tbl:devconv}) converges slowly towards the final value of 3.0 eV.
The cluster calculations exhibit a similar slow convergence behavior.
Interestingly, the cluster calculations also yield excitation energies that are closer to the converged value for the smaller regions (QM2 and QM3) than the QM/MM calculations, which is indicative of a low-quality embedding potential.
Using instead the PE model to represent the environment (PE-QM in Table~\ref{tbl:devconv}), the excitation energies based on any of the quantum regions yield essentially converged results.
This demonstrates nicely that using the PE model enables much smaller quantum regions compared to standard QM/MM without compromising accuracy.

\subsection{Basis Set in the Quantum Region}
\label{sec:basis}
Choosing a basis set for the quantum region represents a major challenge in quantum--classical approaches and thus also for PE-QM calculations because of the risk of electron spill-out, which is the effect where the electron density is pulled into the environment because of too favorable interactions with positively charged areas in the environment.
This can happen because of the lack of exchange repulsion between the quantum and classical regions.
It is therefore not advisable to blindly use the well-known strategy of increasing the size of the basis set until convergence.
Furthermore, care must be taken when following traditional basis-set advice because these are based on investigations performed in vacuum.
For example, the requirement to add diffuse functions when calculating certain properties, such as polarizabilities and hyperpolarizabilities, or when examining excited states, as well as calculations on anionic compounds, may not be valid because such functions extend far into the environment, thus increasing the risk of electron spill-out.
The electron spill-out affects properties in different ways and unfortunately it is not always obvious when there is a problem or how severe it is.
A basis-set analysis should therefore be performed for different sizes of the quantum region if possible so that exchange repulsion is included for the nearest molecules or fragments.
We refer the reader to Refs.~\citenum{Naabo2016Embedding}, \citenum{Reinholdt2017Polarizable}, and \citenum{Fahleson2018} for some examples of the electron spill-out effect on different properties.

\begin{figure}[h!]
    \centering
    \includegraphics[width=0.8\textwidth]{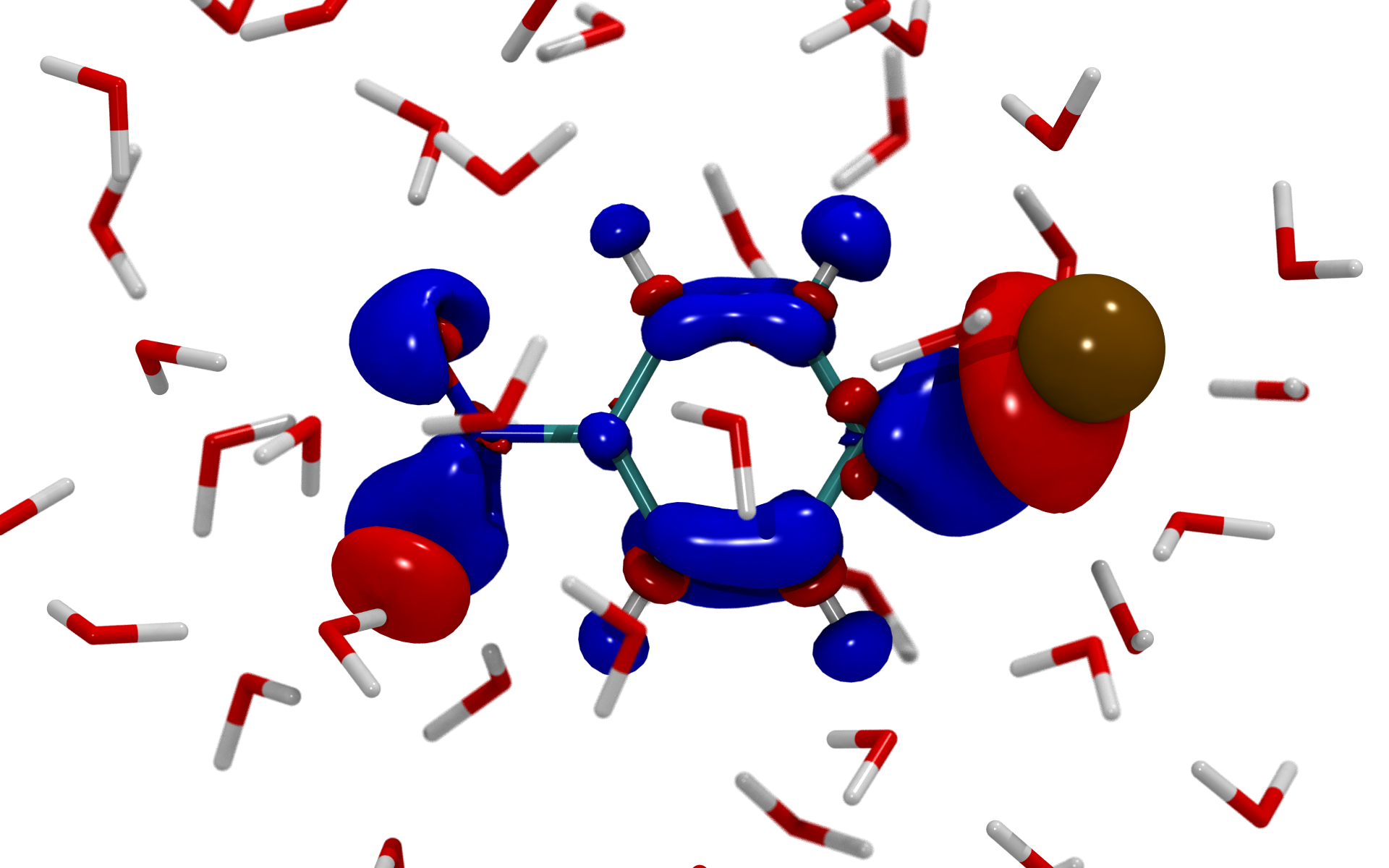}
    \caption{Difference between the electron density of 4-nitrophenolate where the interactions with the solvent environment include or exclude exchange repulsion. The red surface corresponds to areas where there is an excess electron density of 0.001 e$^{\text{-}}$/a$_{0}$. Densities were calculated at CAM-B3LYP/6-31+G* level.}
    \label{fgr:density_difference}
\end{figure}

A concrete example of the electron spill-out effect is shown in Figure~\ref{fgr:density_difference} where a 4-nitrophenolate anion is embedded in a water solvent that includes a sodium cation.
In such situations, it is advisable to include some of the key parts of the environment in the quantum region and to perform a very careful basis-set analysis.
The alternative is to use other more advanced approaches that include exchange repulsion.
Unfortunately, there are usually high computational costs associated with such methods.
We have developed an embedding scheme, polarizable density embedding (PDE)\cite{Olsen2015Polarizable}, that includes exchange-repulsion effects in addition to improved account of the electrostatic interactions.
One of the key features of the new model is efficiency which is achieved by including polarization effects through precomputed polarizabilities, just as in the PE model.
The PDE model will be available in a future release of the PE library.
See Ref.~\citenum{Reinholdt2017Polarizable} for a discussion of the PE and PDE models in relation to electron spill-out effects.

\subsection{Quality of the Embedding Potential}
\label{sec:embedding}
It is usually the representation of the quantum region that tends to dominate the discussion surrounding quantum--classical calculations.
However, it is equally important to consider the description of the classical region, which is often overlooked.
One of the most important aspects related to the environment is the quality of the embedding potential.
Using a high-quality potential is essential to achieve good accuracy of the calculated properties and it is the key point that allows use of small quantum regions (as demonstrated in Section~\ref{sec:qmregionsize}).
The embedding parameters, i.e.\ the permanent multipoles and the dipole--dipole polarizabilities, used in the PE model are usually derived from fragment-based calculations and the key factor is thus the level of theory used for these calculations.

The direct measure of the quality of the embedding potential is the accuracy of the ESP generated by the embedding parameters.
Several studies have shown that use of the aug-cc-pVTZ basis set produces essentially converged ESPs when using either HF or DFT\cite{Soderhjelm2011Conformational,Olsen2015Accuracy,Beerepoot2016Averaged}.
However, the error introduced by the smaller aug-cc-pVDZ basis set has been found to be small overall.\cite{Olsen2015Accuracy,Beerepoot2016Averaged}
Depending on the required accuracy, using the 6-31+G* basis set can also be good compromise between accuracy and computational cost\cite{Olsen2015Accuracy}.
It is clear that using multipoles and polarizabilities will introduce an error compared to the quantum-mechanical ESP.
However, the error is rather constant when comparing the ESP generated by the embedding parameters to the quantum-mechanical ESP when both are derived using the same method, i.e.\ the error is independent of the method used to derive them\cite{Olsen2015Accuracy,Nrby2016Multipole}.
Therefore it makes sense to choose a level of theory for the fragments in the environment that matches the one used for the quantum region, or possibly lower in order to save computational costs.
It is common practice to use a smaller basis set for parts of a system that are not directly of interest.
In a PE-QM calculation, it is also trivial to use a different method for the fragments in the environment, e.g.\ using HF or DFT to generate the embedding potential for use in a PE-CC calculation.

\begin{figure}[h!]
\centering
\includegraphics[width=0.8\textwidth]{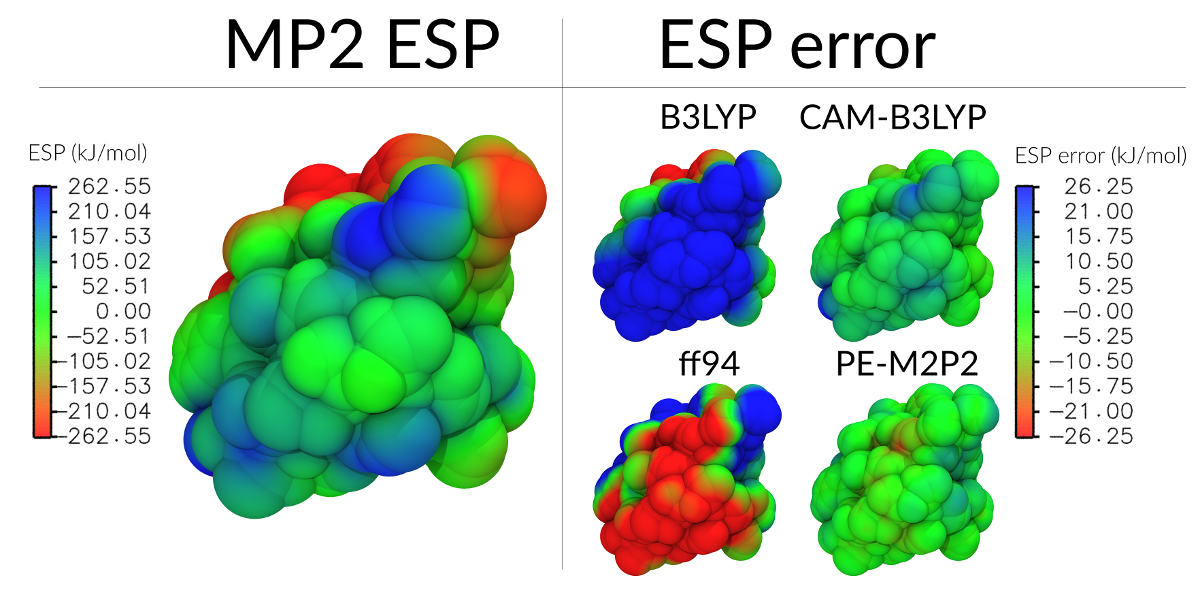}
\caption{Electrostatic potential of insulin on a surface defined by twice the vdW radii. The left panel shows the ESP derived from a MP2 calculation\cite{Jakobsen2013}. The right panel shows the deviation of the ESPs generated by B3LYP, CAM-B3LYP, charges from the AMBER ff94 force field, or the fragment-based M2P2 potential\cite{Olsen2015Accuracy}. The cc-pVDZ basis set was used in all relevant cases.}
\label{fgr:insulin_esp}
\end{figure}

We have shown in several studies that an embedding potential built from fragment-based calculations of atom-centered multipoles truncated at second order (quadrupoles) and atom-centered dipole--dipole polarizabilities derived using the LoProp procedure\cite{Gagliardi2004Local} provides a highly accurate representation both for solvents\cite{olsen2010,Schwabe2011Solvation,Beerepoot2016Averaged}, proteins\cite{Olsen2015Accuracy}, DNA\cite{Norby2016Computational}, and lipids\cite{Witzke2017Averaged}.
An example of the accuracy that can be obtained with this type of embedding potential (often denoted M2P2) is shown in Figure~\ref{fgr:insulin_esp}.
In this figure, we compare ESPs derived from B3LYP, CAM-B3LYP, AMBER ff94 point charges, or the M2P2 potential to the ESP from a full MP2 calculation on insulin.
It can be seen that the M2P2 potential (using B3LYP in the fragment calculations) has very small errors of around 5 kJ/mol which is comparable to the ESP from a full CAM-B3LYP calculation.
Furthermore, it is also apparent that point charges from standard force fields, here exemplified by the AMBER ff94 force field, cannot reproduce the ESP very well.
With such a poor reproduction of the ESP it is necessary to use much larger quantum regions compared to the requirements when using the high-quality M2P2 potential.
Note that the LoProp procedure requires an ANO-type basis set\cite{Gagliardi2004Local}, several of which are available in Dalton (they can be identified by the \texttt{loprop-} prefix).
The M2P2 potential can be readily obtained using the PyFraME package based on either Dalton calculations and the LoProp for Dalton script\cite{Vahtras2014LoProp} or MOLCAS.
See Section~\ref{ssec:generatefiles} for details on how to generate the potentials.
Finally, we stress that LoProp-based multipoles should never be truncated at first order (i.e.\ charges and dipoles only) because of the convergence behavior\cite{Olsen2015Accuracy,Nrby2016Multipole}.

\subsection{Size of the Environment}
\label{sec:mmregionsize}
Here we discuss the question of how much of the environment should be included.
This is also a balance between accuracy and computational cost although it is usually not as critical as it is with respect to the quantum region.
For homogeneous systems, such as solute-solvent systems, it is usually sufficient to include solvent molecules within 10 to 15 \AA{} from the quantum region to obtain reasonably converged properties\cite{Steinmann2014Nuclear,Beerepoot2014Convergence,Norby2017Modeling}.
The situation is rather different in heterogeneous systems such as proteins where the entire protein and preferably also a solvent shell around the protein is usually included in the classical region.
In both cases, it is advisable to perform an analysis to make sure the property of interest is converged.
Such analysis can be performed based on a single molecular structure or on a series of structures.
Using a single structure is a more strict requirement so faster convergence is typically observed for statistically averaged properties.
See Section~\ref{sec:sampling} for more details concerning statistical sampling.

\begin{figure}[h!]
  \centering
  \includegraphics[width=0.85\textwidth]{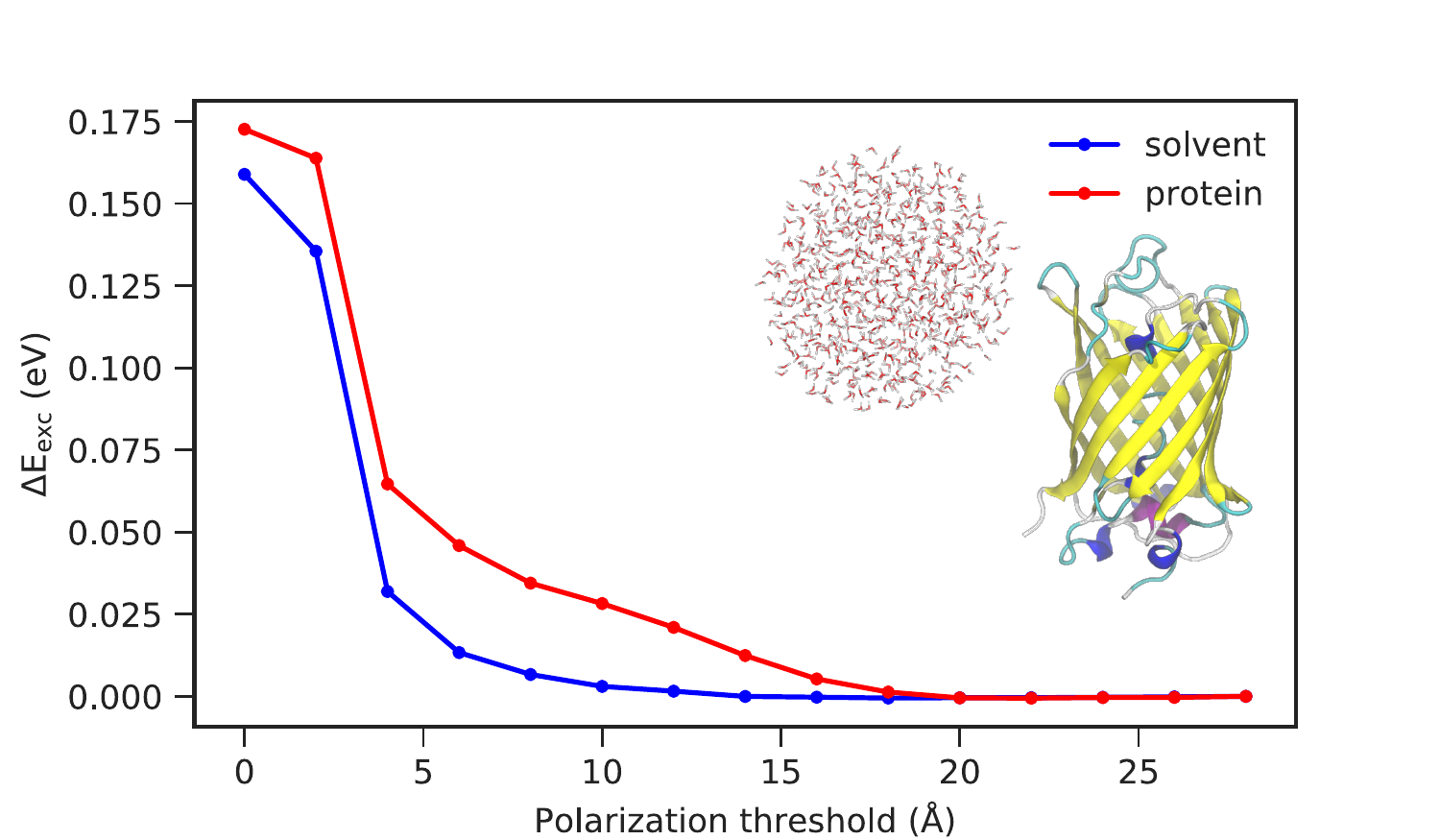}
  \caption{Deviation in excitation energy, $\Delta E$, due to removal of polarizabilities outside a polarization threshold of the HBDI chromophore when embedded either in water (blue) or in the GFP (red).
  The deviation is taken with respect to a reference value obtained at a polarization threshold of 20 \AA.
  Reproduced with data from Ref.~\citenum{Beerepoot2014Convergence}.}
  \label{fgr:ss}
\end{figure}

As discussed earlier, the most computationally intensive part associated with the environment is the self-consistent determination of induced dipoles.
It might therefore be tempting to use different cutoff distances for the electrostatic and polarization interactions, considering the lower range of the latter, in order to save computational costs.
Indeed this has been investigated in several studies\cite{Osted2004Linear,Osted2006Statistical,Sderhjelm2009Protein,Soderhjelm2009Calculation,Curutchet2012Energy}.
However, \citeauthor{Beerepoot2014Convergence}\cite{Beerepoot2014Convergence} found that such an approach is not necessary to get a good balance between accuracy and computational efficiency in a PE-QM calculation.
Figure~\ref{fgr:ss}, which is based on data from Ref.~\citenum{Beerepoot2014Convergence}, shows the convergence of the lowest intense transition in the anionic form of the HBDI chromophore embedded in the solvated GFP protein or in water solution, with respect to a polarization cutoff threshold.
The permanent electrostatics are included up to 30 \AA{} away from the chromophore in both cases.
The polarization interactions in the solvent environment converge at around 10 \AA{} which is so close to the full convergence threshold (10--15 \AA{}) that the computational gain is negligible.
The effects of polarization extend much farther in the protein environment ($~$20 \AA{}) due to the charged residues which, even though they may be far away from the chromophore, have a significant effect via the induced dipoles.
However, the computational speed-up gained by using the converged polarization cutoff threshold was only $~$10\% which is not worthwhile.

So far we have only considered the cost associated with a PE-QM calculation itself.
There is of course an additional computational overhead associated with large environments which lies in the computation of fragment multipoles and polarizabilities.
In cases where the computational cost is a bottleneck, it can be beneficial to split the environment into two parts, e.g.\ based on a distance threshold\cite{Beerepoot2016Averaged}.
Fragments that are close to the quantum region are treated as before, i.e.\ the embedding parameters are determined based on fragment calculations, whereas averaged parameters can be used for fragments that are farther away from the quantum region.
Such averaged parameters exist for a series of solvents\cite{Beerepoot2016Averaged}, lipids\cite{Witzke2017Averaged}, and nucleotides\cite{Norby2016Computational}, or, alternatively, the parameters can also be extracted from standard molecular-mechanics force fields.
To determine a suitable distance threshold it is necessary to perform benchmark calculations (see Ref.~\citenum{Beerepoot2016Averaged} for examples).

\subsection{Quantum--Classical Interface}
\label{sec:qmmminterface}
\begin{figure}[h!]
\centering
\includegraphics[width=0.7\textwidth]{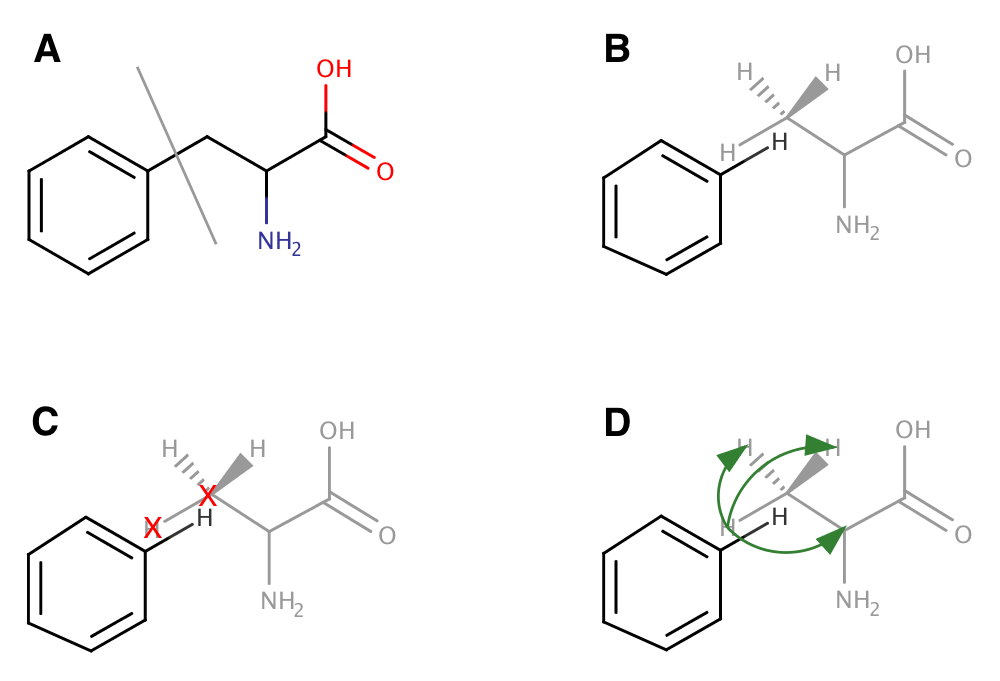}
\caption{Illustration of different ways to treat the quantum--classical interface. Panel A shows the full system before the bond is cut. Panel B shows the system after the bond is cut and the open valences are saturated with hydrogens. The quantum region is shown in black and the classical region is gray. Panels C and D illustrate the deletion and redistribution of parameters, respectively.}
\label{fgr:border}
\end{figure}

In some cases it is necessary to let the border between the quantum and classical regions cross a covalent bond, for example when modeling properties of a covalently bonded chromophore in a protein.
To handle such cases, we rely on a simple hydrogen link-atom approach.
Special attention should be given to this quantum--classical interface because of the risk of over-polarization or electron spill-out which can occur if classical sites are too close to the quantum region.
The situation is illustrated in Figure~\ref{fgr:border} using a simple model system consisting of a single phenylalanine amino acid.
In Figure~\ref{fgr:border}A the gray line shows where the bond will be cut thus separating the phenyl group, which will be the quantum region, from the remainder, which is treated classically.
The dangling bonds are saturated with hydrogen atoms (called link-atoms) as shown in Figure~\ref{fgr:border}B.
The role of the link-atoms is to mimic the effects from the carbons they replaced.
The embedding parameters of the classical fragment are derived including the hydrogen link-atom and the quantum link-atom is preserved in the PE-QM calculation.
This results in classical atoms that are basically overlapping with quantum atoms and it is therefore necessary to modify the classical link-atom and the classical carbon bonded to it (as a minimum).
There are two recommended ways to deal with these atoms: (1) all embedding parameters are removed (illustrated in Figure~\ref{fgr:border}C) or (2) charges are redistributed, to the nearest one, two, or three sites, or to all sites, while all other parameters are removed (illustrated in Figure~\ref{fgr:border}D for the classical link-atom).
The latter preserves the overall charge of the system.
It may also be necessary to remove or redistribute charges of other nearby atoms to avoid over-polarization issues.
For example, in the illustration there are the two hydrogens that are close to the quantum link-atom.
Due to this proximity, the electric field from the electron density that surrounds the quantum link-atom could induce large dipoles on the hydrogens causing a positive feedback loop.
Alternatively, such issues could possibly be alleviated by damping the electric fields from the quantum system (using the \texttt{.DAMP CORE} keyword). 

To activate removal or redistribution of embedding parameters close to the quantum region, the \texttt{.BORDER} keyword has to be included in the Dalton input file under the \texttt{*PEQM} section.
The different removal/redistribution schemes described above are controlled by three different options belonging under the \texttt{.BORDER} keyword.
The removal scheme, illustrated in Figure~\ref{fgr:border}C, is used by adding \texttt{REMOVE} followed by a distance and the unit of the distance (\texttt{AU} or \texttt{AA}).
For example, \texttt{REMOVE 0.5 AA} which would remove all parameters within 0.5 \AA{} of an atom in the quantum region.
A short distance, such as 0.5 \AA{}, is usually adequate to take care of the overlapping atoms.
To remove additional atoms it is necessary to measure the distances to make sure that only the intended atoms are affected.
The redistribution scheme, illustrated in Figure~\ref{fgr:border}D, is split into two different option.
The first is \texttt{REDIST} which requires also that the order up to which multipoles are redistributed is specified (the sign determines whether polarizabilities are also redistributed), as well as the distance, unit of the distance, and finally also the number of sites to which the parameters will be distributed to.
This could for example be \texttt{REDIST 1 0.5 AA 3} which specifies that charges within 0.5 \AA{} are redistributed to three nearest sites and that all other parameters are removed.
The second redistribution scheme, which is the redistribution of nearby charges to all other sites and removal of other parameters, is used by adding \texttt{CHGRED}, a distance, and unit of the distance, e.g.\ \texttt{CHGRED 0.5 AA} which redistributes the charge within 0.5 \AA{} to all other sites.

\begin{figure}[h!]
\centering
\includegraphics[width=0.85\textwidth]{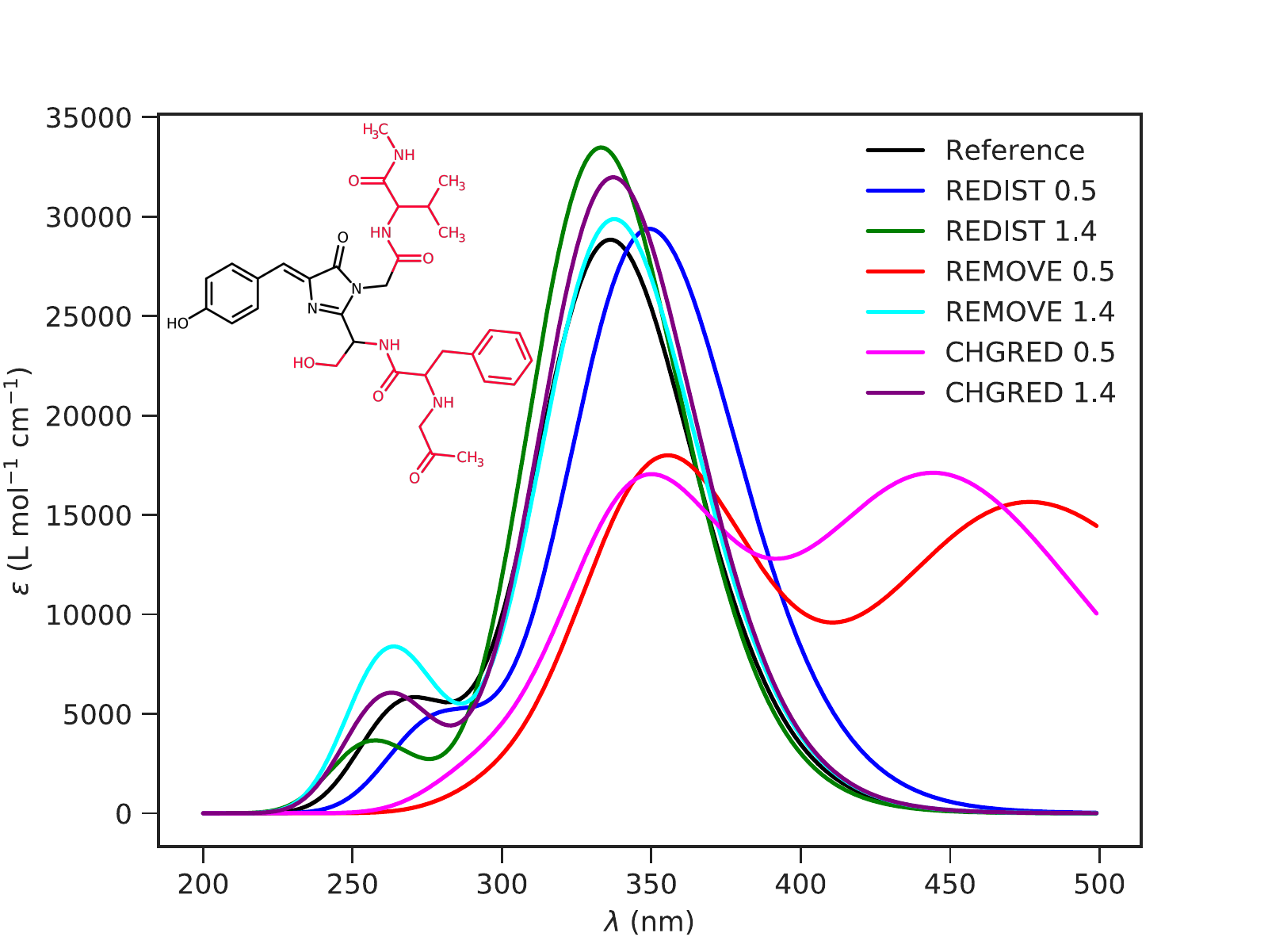}
\caption{UV/Vis spectra showing the effect of the different ways to treat the quantum-classical interface. The spectra are based on CAM-B3LYP/6-31+G* calculations on a model system containing the HBDI chromophore as shown in the inset (black parts were in the quantum region while the red parts were treated classically). The reference is a full quantum-mechanical calculation on the whole model system.}
\label{fgr:quantum-classical}
\end{figure}

One way to determine the optimal procedure for the quantum--classical interface is to perform calculations on a small model system where a full quantum-mechanical calculation is possible.
An example of such a strategy is illustrated in Figure~\ref{fgr:quantum-classical} for a UV/Vis spectrum of the HBDI chromophore where both redistribution and removal of charges was examined.
A distance of 0.5 \AA{} will affect the classical link-atom and the carbon it is attached to, whereas the 1.4 \AA{} distance will additionally affect any hydrogens attached to the said carbon.
The example shows that the different schemes can have drastically different effects on the properties of the quantum region which underlines the importance of performing such investigations.
Furthermore, it is clear that two of the investigated choices are inadequate yielding qualitatively wrong spectra, whereas the others result in rather faithful reproductions of the reference spectrum.
Unfortunately, it is not possible to generalize from such an investigation on a particular system.
The optimal procedure must be determined on a case-by-case basis.
That said, it is appealing to choose a scheme that preserves the overall charge of the system though it does not necessarily mean that it is the most favorable which is also evident from the example.
Furthermore, a high sensitivity to the treatment of the quantum--classical interface may be an indication of a too small quantum region.

\subsection{Local Field Effects}
\label{sec:eef}
In the presence of external electric fields, which is relevant when considering spectroscopies of embedded molecules, the local field acting on the embedded molecule is generally modified by the presence of the polarizable environment.
The modification of the local field is due to the direct polarization of the environment by the external field.
Hence, the local field acting on the embedded molecule will in general be different from the externally applied field.
Such field modifications have traditionally been accounted for by introducing local field factors\cite{Onsager1936}.
Within the PE approach, local field effects enter into the definitions of response properties of the embedded molecule as derived and discussed in Ref.~\citenum{List2016Local}, where they are termed effective external field (EEF) effects.
Only the strength of the field is modified, and not the frequency.
This means that excitation energies are not affected by the EEF but properties like transition moments and polarizabilities will change.
The importance of including the EEF is clearly seen when comparing results of response property calculations based on a full quantum mechanical approach to the corresponding PE-QM calculations.
The results in Table~\ref{tbl:LF} refer to calculations on a model system of the DsRed fluorescent protein consisting of 242 atoms that are from Ref.\citenum{List2016Local}, and we refer to this reference for further details regarding the calculations.
As seen from the table, the PE model predicts the position of the lowest singlet excitation energy in very good agreement with full quantum-mechanical calculations and, as expected, the EEF does not change the result of this excitation energy.
On the other hand, including the EEF in the calculation of the response properties, here the oscillator strength and the two-photon absorption (2PA) cross section, leads to results that are in much better agreement with the full quantum-mechanical results.
It is therefore highly recommended to include the EEF in PE-QM calculations of electrical response properties.
The EEF is activated by adding the \texttt{.EEF} keyword to the \texttt{*PEQM} section in the Dalton input file.

\begin{table}[h!]
    \centering
    \caption{Excitation energies (in eV), oscillator strengths and 2PA cross sections (in GM) of the lowest singlet excited state of a cluster model of the DsRed protein (242 atoms). Results from Ref.~\citenum{List2016Local}.}
    \label{tbl:LF}
    \begin{threeparttable}
    \begin{tabular}{lccc}
	\headrow
    \thead{Method} & \thead{$\bm{\Delta}$E} & \thead{f} & \thead{$\bm{\sigma}$} \\
    PE      &  2.96 &  1.11 &  77 \\
    PE(EEF) &  2.96 &  0.80 &  42 \\
    Full QM &  2.92 &  0.84 &  38 \\
    \hline
    \end{tabular}
    \end{threeparttable}
\end{table}

\subsection{Statistical Sampling}
\label{sec:sampling}
To obtain a realistic estimate of a response property it is necessary to include statistical sampling.
The sampling is often obtained through MD simulations from which a number of snapshots are extracted.
The property of interest is then obtained as an average based on individual calculations on each of the snapshots.
This procedure raises two important questions.
First, how to obtain molecular structures that are adequate to use in response property calculations, and, second, how many snapshots are needed to obtain converged properties?

Most of the common force fields used in classical molecular-mechanics-based MD are focused on reproducing bulk properties rather than microscopic properties which can be very sensitive to the molecular structure.
Moreover, it is necessary to rely on generalized force fields when there are no optimized parameters available, which is often the case for the parts that will be in the quantum region.
For this reason, the extracted geometries can be sub-optimal, in particular for the quantum region.
Using instead QM/MM MD methods is an attractive option, but the timescales accessible with such methods can be too limited for many systems.
An alternative is to refine the geometry of the snapshots from the classical MD before using them in the final response property calculation.
This may be done by optimizing the geometry of the quantum region using QM/MM methods while keeping the environment frozen.
Obviously, this has a negative effect on the dynamics of the quantum region, but this is sometimes more preferable than a poor geometry.

\begin{figure}[h!]
  \centering
  \includegraphics[width=0.85\textwidth]{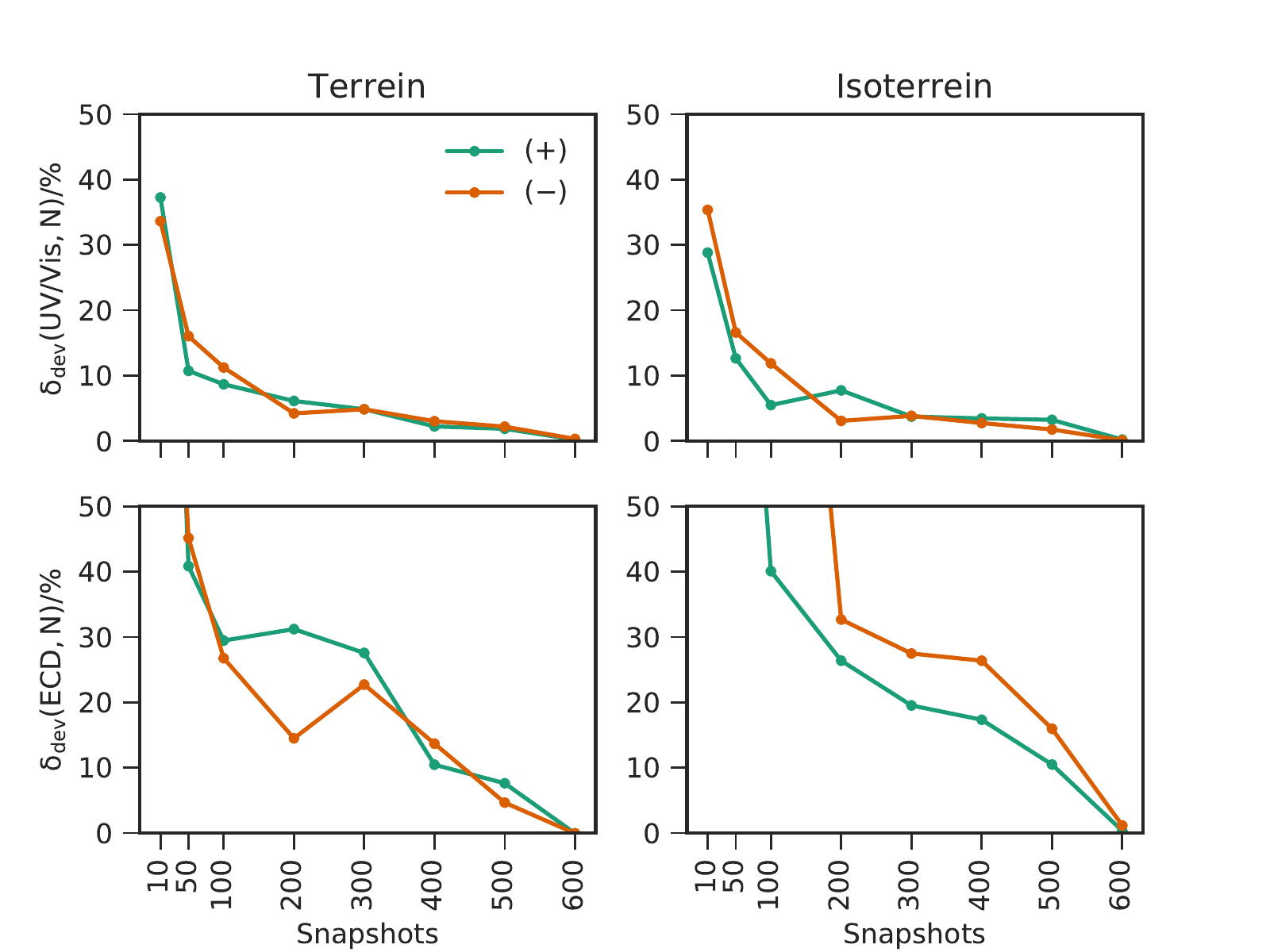}
  \caption{Convergence of UV/Vis (top) and ECD (bottom) spectra of (+/-)-terrein and (+/-)-isoterrein with respect to the number of configurations.
 Reproduced with data from Ref.~\citenum{Norby2017Modeling}.}
  \label{fgr:convergence_terrein}
\end{figure}

Concerning the convergence of properties, the general recommendation is to investigate the convergence of the particular property with respect to the number of snapshots.
The decisive factors are the rigidity of the molecular structure and the sensitivity of the property to the structural dynamics.
Examples of relatively insensitive properties include the excitation energy and oscillator strengths.
These have rather narrow distributions and usually require only a modest amount of snapshots to reach reasonable convergence, typically around 100 snapshots for solute-solvent system and less for more rigid systems.
Other properties may require a larger number of snapshots, for example, multiphoton absorption strengths~\cite{Steindal2016} and electronic circular dichroism (ECD) \cite{cappelli2016integrated,Norby2017Modeling}.
The requirements of the latter is illustrated in Figure~\ref{fgr:convergence_terrein}, where we see, in the same system and for the same geometries, a much slower convergence for the ECD than for the UV/Vis spectrum.

\subsection{Implicit Solvation}
\label{sec:implexpl}
Finally, we also mention the option to model bulk solvent effects implicitly through the FixSol\cite{Thellamurege2012FixSol} conductor-like solvation model\cite{Norby2016Computational} which will be available in an upcoming release of the PE library.
Implicit solvation is activated by adding the \texttt{.SOLVATION} keyword under the \texttt{*PEQM} section in the Dalton input file.
This will use the FIXPVA2\cite{Thellamurege2012FixSol} algorithm to generate the cavity and by default use a water solvent but other solvents can be specified as an additional keyword under \texttt{.SOLVATION}.

\begin{figure}[h!]
\centering
\includegraphics[width=0.85\textwidth]{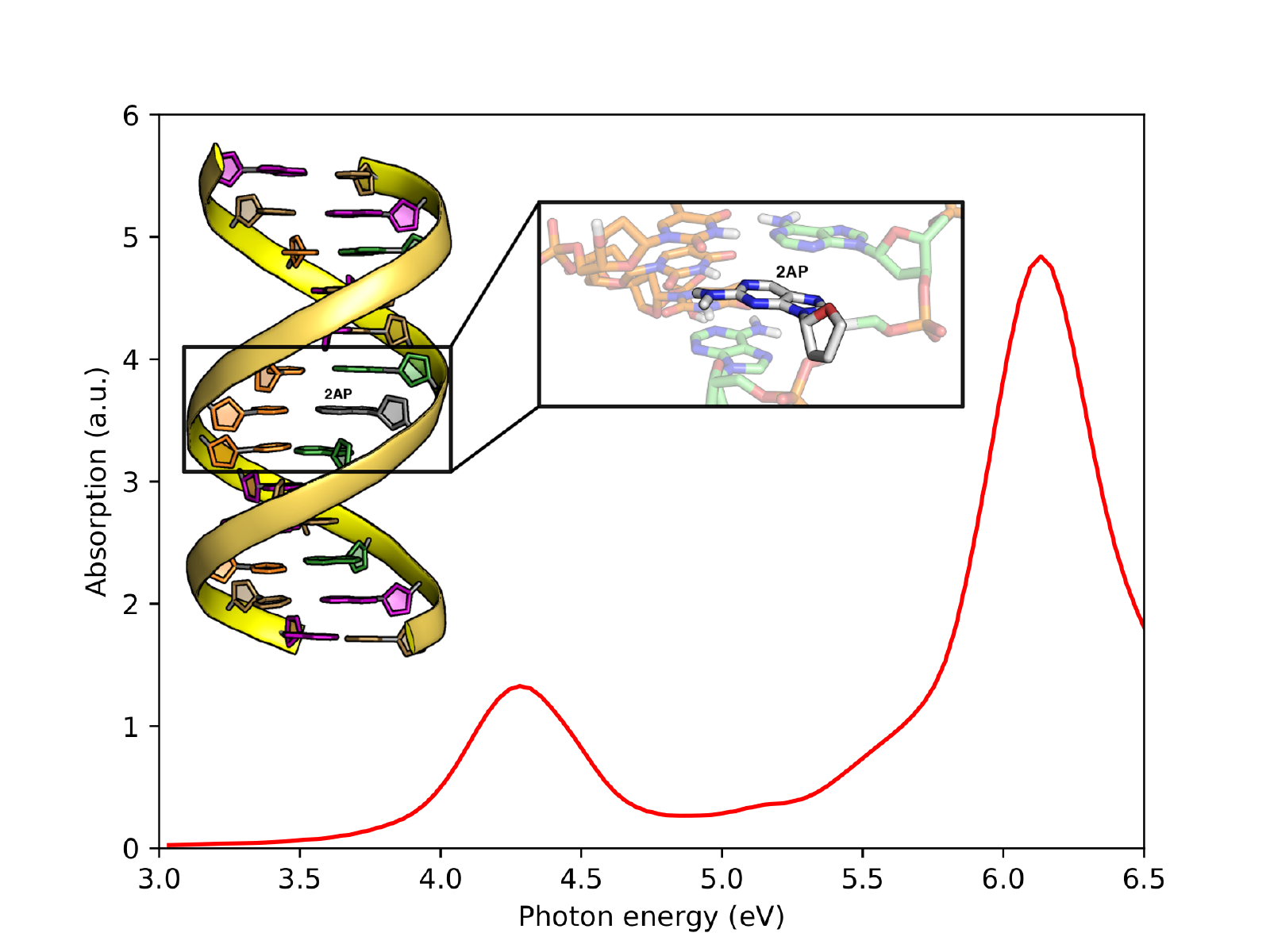}
\caption{UV/Vis spectrum and molecular illustration of 2-aminopurine (2AP) embedded in a DNA double-helix with implicit solvent. Reproduced with data from Ref.~\citenum{pedersen2014damped}.}
\label{fgr:fixsol}
\end{figure}

Using implicit solvation is particularly relevant when computing properties of large (bio)molecules where inclusion of bulk solvation effects are relevant.
To illustrate, we show the UV/Vis spectrum of the 2-aminopurine optical probe embedded in a solvated DNA double-helix in Figure~\ref{fgr:fixsol}.
The 2-aminopurine fragment is placed in quantum region, which is treated at CAM-B3LYP/6-31+G* level, while the remaining fragments in the double-helix are modeled using the PE model and the water solvent is included via FixSol.
The advantage of the continuum solvation is that it includes statistical sampling implicitly and it is therefore only necessary to sample conformations of the DNA helical structure.

\section{Summary}
In this tutorial, we have described and discussed practical aspects related to the use of the PE model in calculations of response properties of molecules embedded in atomistic environments such as solvents or proteins.
We have demonstrated how the input for a response-property calculation can be prepared starting from a molecular structure and we have given examples that illustrate how to run the calculations in practice.
Calculations involving embedded molecules are much more involved compared to the equivalent calculation on molecules in vacuum.
There are many additional complications, which, if handled poorly, can easily compromise the accuracy of the results.
To make it easier for new users, we have provided guidelines concerning several aspects that are important to consider in order to be able to efficiently compute accurate response properties.
This included a discussion of the requirements with respect to the size of the quantum region and the choice of basis set.
Best practices concerning the treatment of the environment have also been presented, which covered considerations concerning the size of the environment, the quality of the potential describing the environment, treatment of quantum--classical interfaces, and local field effects.
We have also touched upon issues related to statistical sampling and the possibility to use implicit solvation to model bulk solvent effects on large molecules such as proteins or DNA.
This tutorial can thus be used to guide new users through the otherwise complicated process of going from molecular structure to response property.

\section*{Acknowledgements}
Computation/simulation for the work described in this paper was supported by the DeIC National HPC Centre, SDU.
C.\ S.\ thanks the Danish Council for Independent Research for financial support (Grant ID: DFF--4181-00370).
J.\ K.\ acknowledges financial support from
the Villum Foundation, the Danish Council for
Independent Research (Grant ID: DFF--7014-00050B), and from the H2020-MSCA-ITN-2017 COSINE Training network for COmputational Spectroscopy In Natural sciences and Engineering (Project ID: 765739).
J.\ M.\ H.\ O.\ thanks the Carlsberg Foundation for financial support (Grant ID: CF15-0823).



\bibliography{main}



\end{document}